\documentclass{svjour3}                     
\smartqed  
\usepackage{graphicx}
\usepackage[sort&compress]{natbib}
\usepackage{color,amsmath,amssymb,amsfonts,array}
\usepackage{varioref}

\bibpunct{(}{)}{;}{a}{}{;}

\newcommand{\tc}{\text{:}} 

\newtheorem{thm}{Theorem}
\newtheorem{lem}[thm]{Lemma} 
\newtheorem{prop}[thm]{Proposition}
\newtheorem{cor}[thm]{Corollary}

\begin{document}


\title{Identifying the Rooted Species Tree from the Distribution of Unrooted Gene Trees under the Coalescent}

\date \today


\titlerunning{Species Tree from Gene Trees}        

\author{Elizabeth S.~Allman\and James H.~Degnan \and  John A.~Rhodes}

\authorrunning{Allman, Degnan, and Rhodes} 

\institute{E. S. Allman \at
              Department of Mathematics and Statistics,
              University of Alaska Fairbanks,\\
              PO Box 756660, Fairbanks, AK 99775 USA\\
               \email{e.allman@alaska.edu}
            \and
            Corresponding author:\\
            J. H. Degnan \at
              Department of Mathematics and Statistics,
              University of Canterbury\\
              Private Bag 4800,
              Christchurch, New Zealand\\
              \email{J.Degnan@math.canterbury.ac.nz}           
           \and
           J. A. Rhodes \at
              Department of Mathematics and Statistics,
              University of Alaska Fairbanks,\\
              PO Box 756660, Fairbanks, AK 99775 USA\\
               \email{j.rhodes@alaska.edu}
}

\date{Received: date / Accepted: date}

\maketitle

\date\today\\

\begin{abstract}

Gene trees are evolutionary trees representing the ancestry of genes sampled from multiple populations.  Species trees represent populations of individuals --- each with many genes --- splitting into new populations or species.  The coalescent process, which models ancestry of gene copies within populations, is often used to model the probability distribution of gene trees given a fixed species tree.  This multispecies coalescent model provides a framework for phylogeneticists to infer species trees from gene trees using maximum likelihood or Bayesian approaches.  Because the coalescent models a branching process over time, all trees are typically assumed to be rooted in this setting.  Often, however, gene trees inferred by traditional phylogenetic methods are unrooted.  

We investigate probabilities of unrooted gene trees under the multispecies coalescent model. We show that when there are four species with one gene sampled per species, the distribution of unrooted gene tree topologies identifies the unrooted species tree topology and some, but not all, information in the species tree edges (branch lengths). The location of the root on the species tree is not identifiable in this situation.  However, for 5 or more species with one gene sampled per species, we show that the distribution of unrooted gene tree topologies identifies the rooted species tree topology and all its internal branch lengths. 
The length of any pendant branch leading to a leaf of the species tree is also identifiable for any species from which more than one gene is sampled.
\end{abstract}

\section{Introduction}

The goal of a phylogenetic study is often to infer an evolutionary tree depicting the history of speciation events that lead to a currently extant set of taxa.  In these \emph{species trees},  speciation events are idealized as populations instantaneously diverging into two populations that no longer exchange genes.  Such trees are often estimated indirectly, from DNA sequences for orthologous genes from the extant species.  A common assumption has been that such an inferred \emph{gene tree} has a high probability of having the same topology as the species tree. Recently, however, increasing attention has been given to population genetic issues that lead to differences between gene and species trees, and how potentially discordant trees for many genes might be utilized in species tree inference.

Methods that infer gene trees, such as maximum likelihood (ML) using standard DNA substitution models, typically can estimate the expected number of mutations on the edges of a tree, but not the direction of time.  
Phylogenetic methods therefore often estimate \emph{unrooted} gene trees.  In many cases, the root of a tree can be inferred by including data on an \emph{outgroup}, \emph{i.e.}, a species believed to be less closely related to the species of interest than any of those are to each other \citep{rokas2003,poe2004,jennings2005}.  However, outgroup species which are too distantly related to the ingroup taxa may lead to unreliable inference, and in some cases appropriate outgroup species are not known \citep{graham2002,huelsenbeck2002}.  The root of a gene tree can alternately be inferred under a \emph{molecular clock} assumption, \emph{i.e.,} if mutation rates are constant throughout the edges of a tree.
In many empirical studies, however, such a molecular clock assumption is violated.  Furthermore, without a molecular clock, inferred branch lengths on gene trees may not directly reflect evolutionary time, as substitution rates vary from branch to branch.
For these reasons, one may have more confidence in the inference of unrooted topological gene trees than in metric and/or rooted versions.

\begin{figure*}[h]\label{F:T15}
\begin{center}
\includegraphics[width=.32\textwidth]{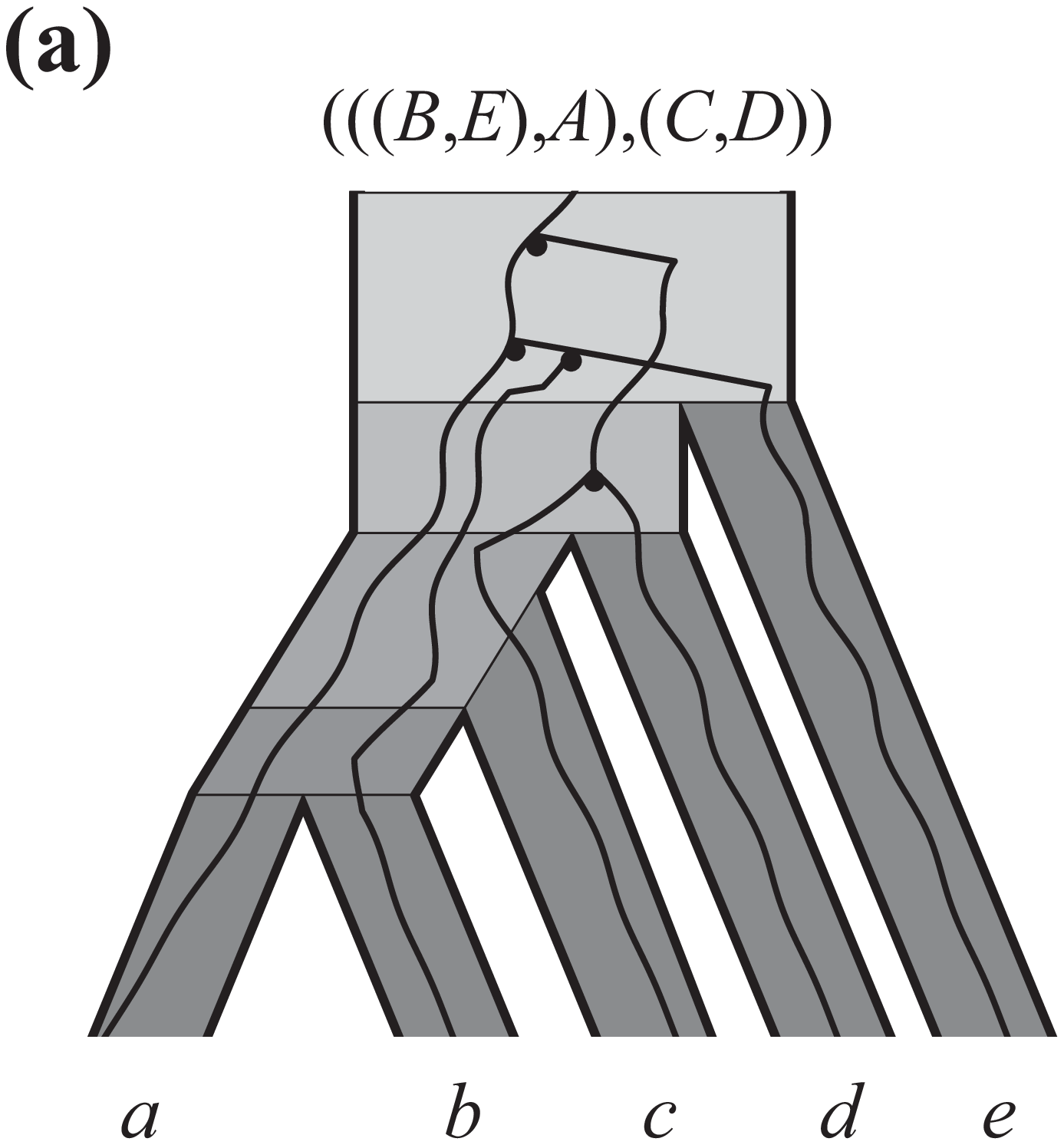}
\includegraphics[width=.32\textwidth]{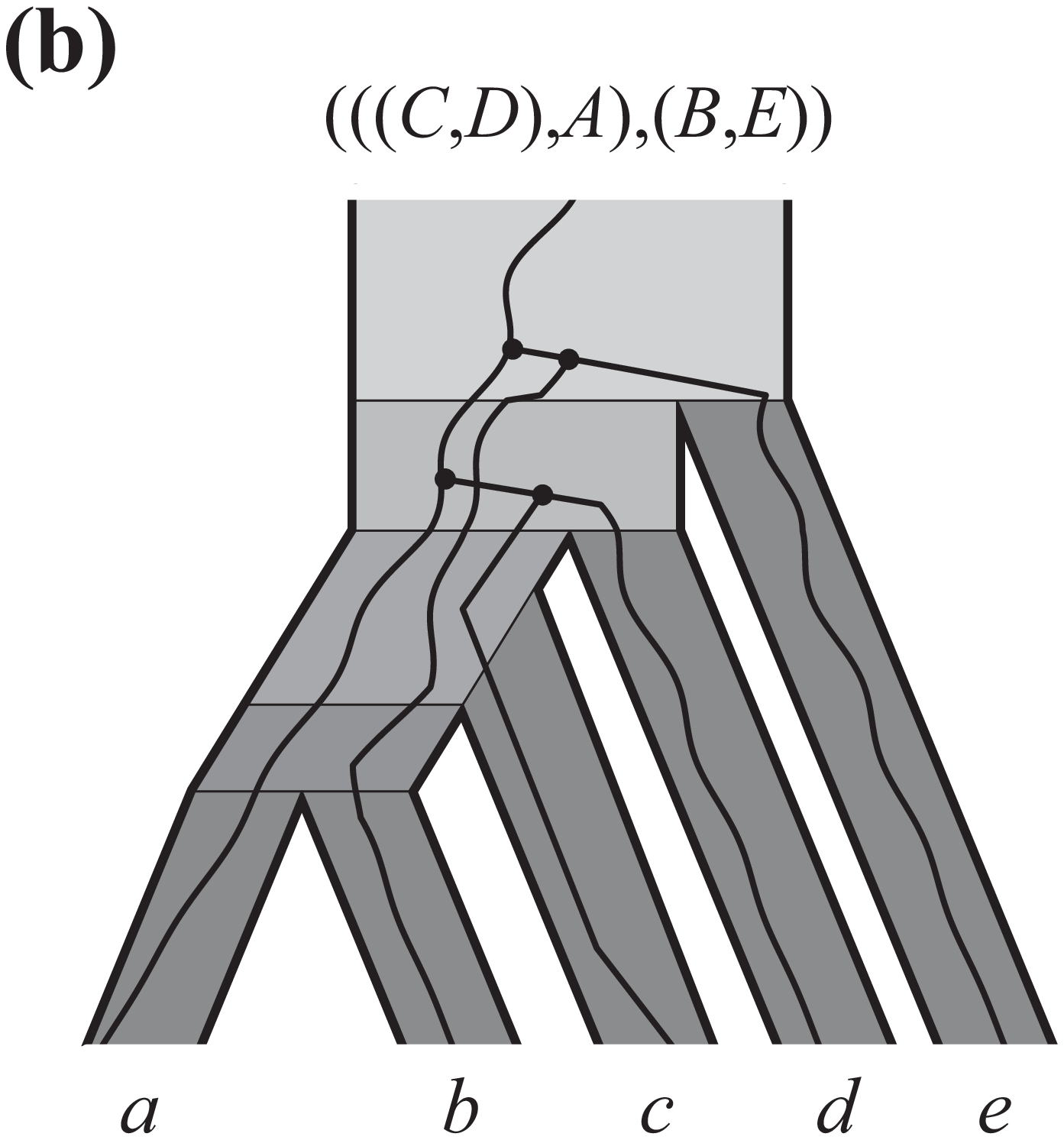}
\includegraphics[width=.32\textwidth]{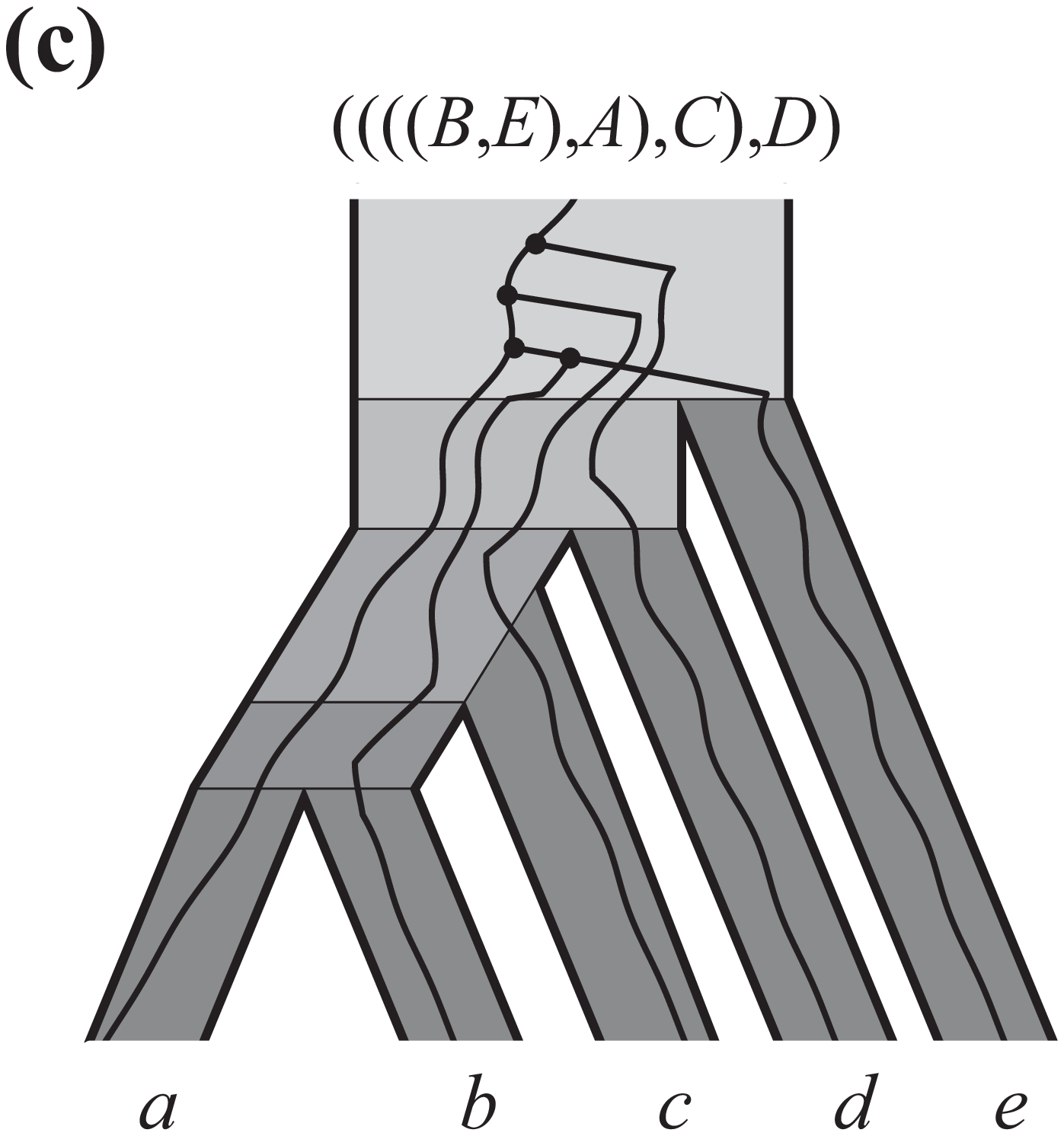}
\vskip 3mm
\includegraphics[width=.32\textwidth]{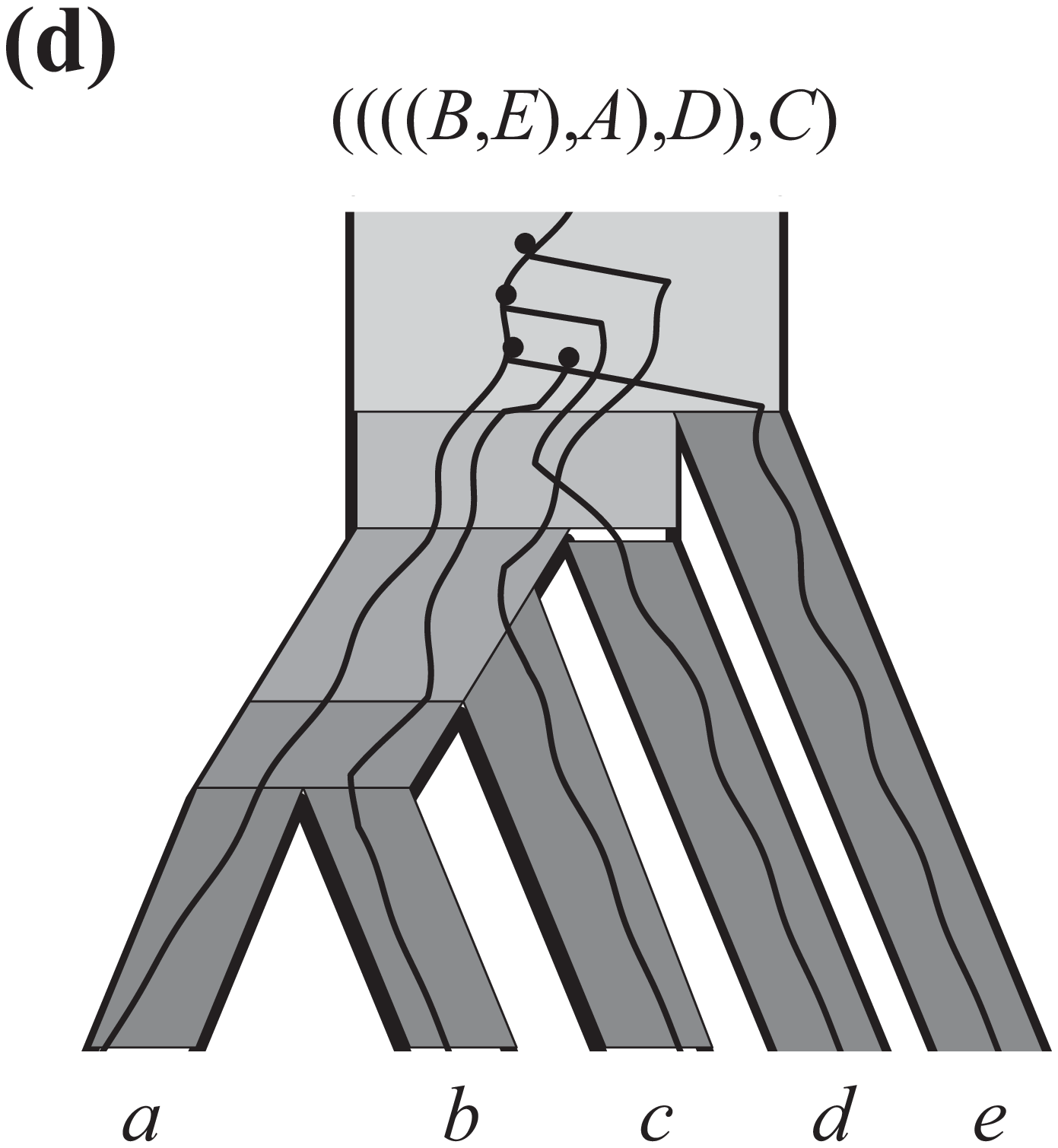}
\includegraphics[width=.32\textwidth]{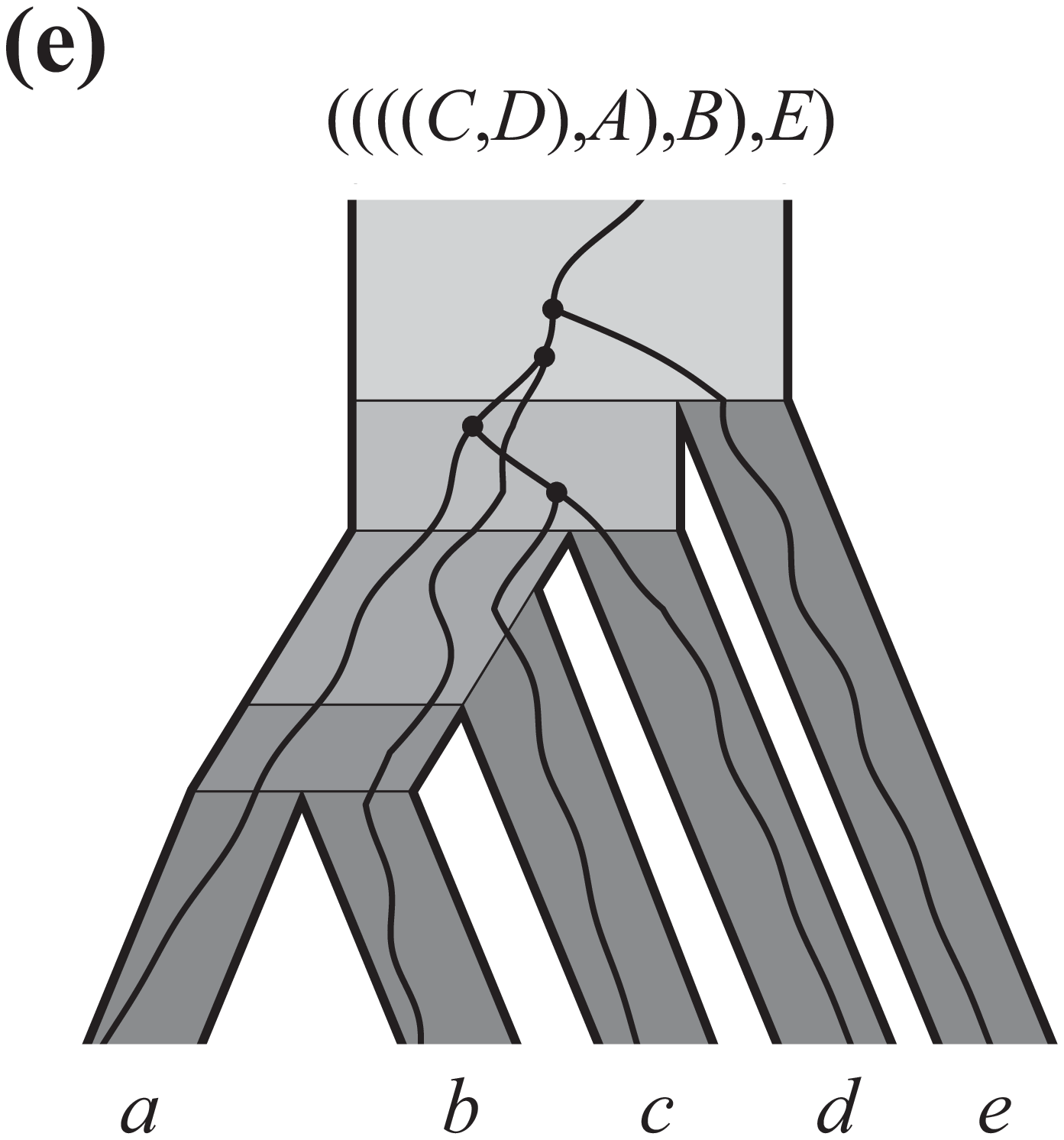}
\includegraphics[width=.32\textwidth]{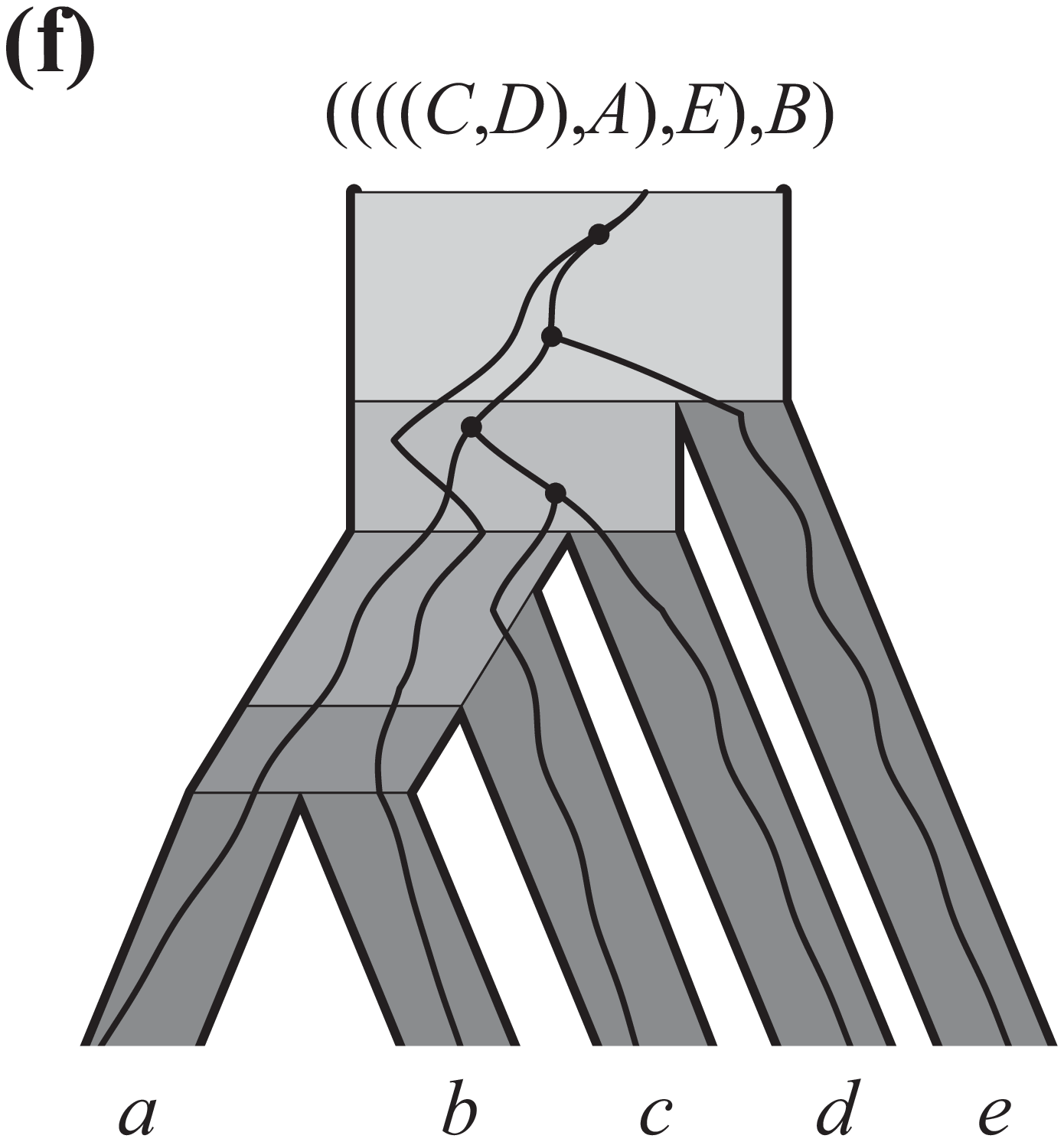}
\vskip 3mm
\includegraphics[width=.32\textwidth]{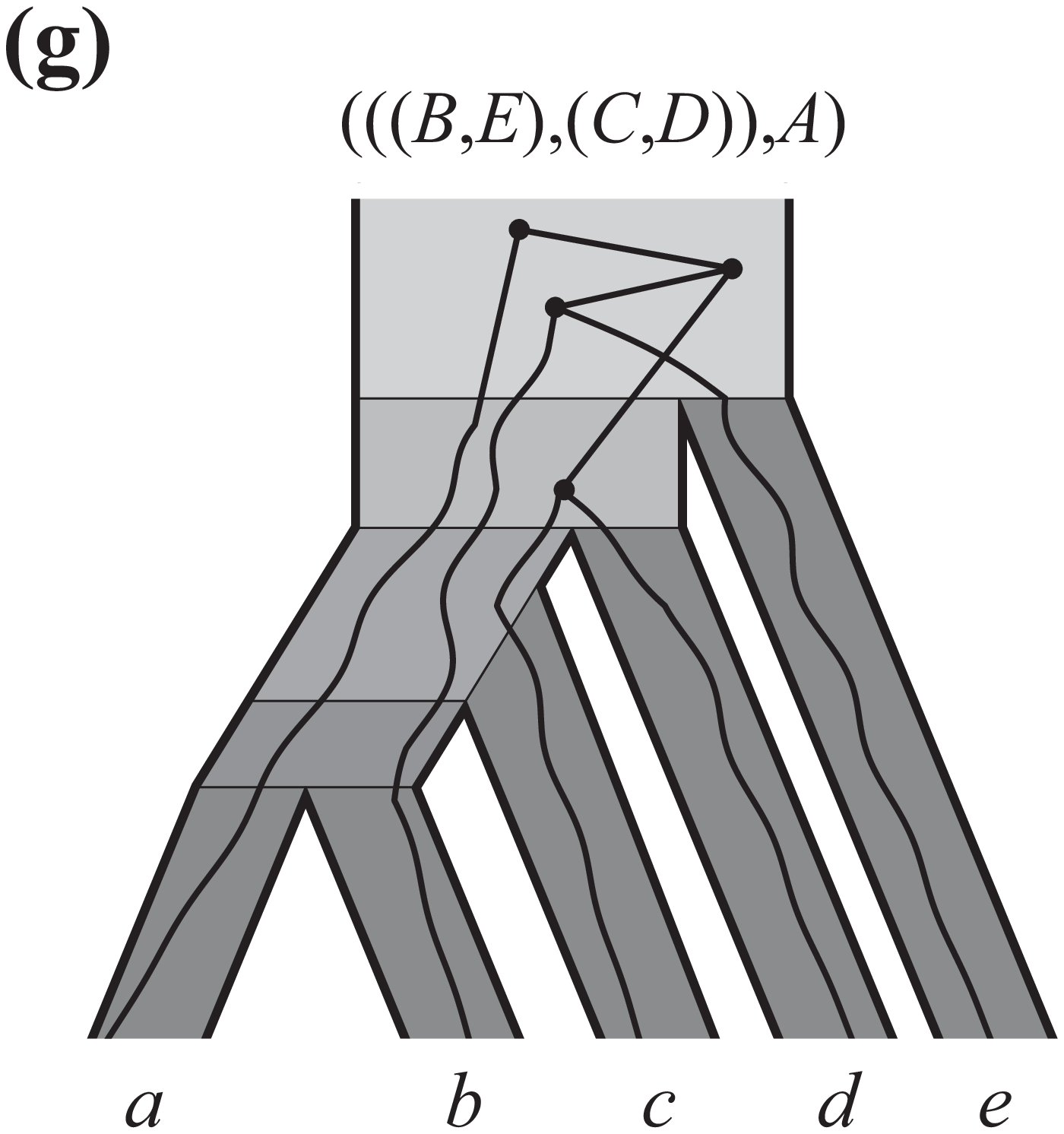}
\includegraphics[width=.32\textwidth]{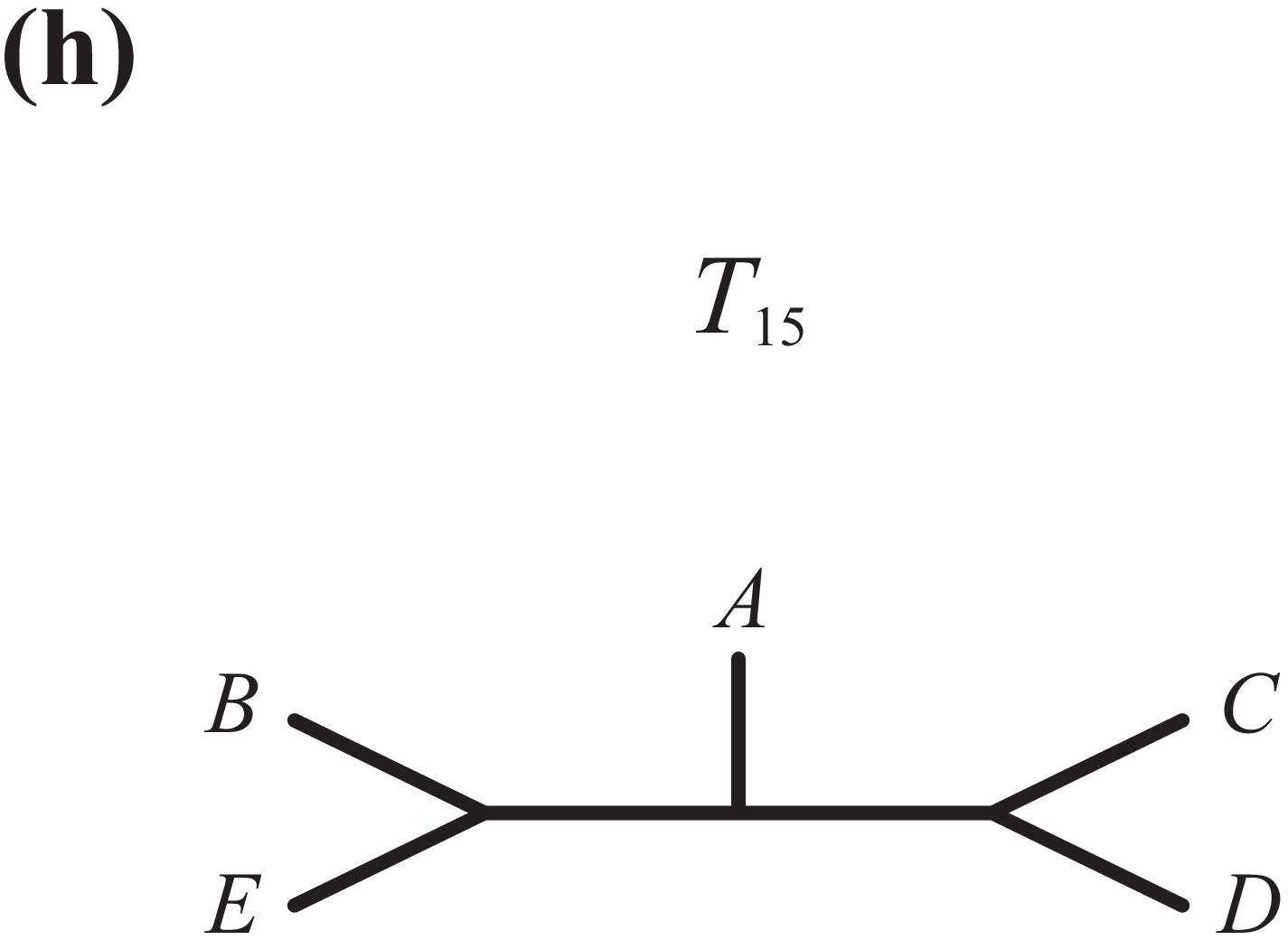}
\end{center}

\caption{The unrooted gene tree $T_{15}$ in the species tree $((((a,b),c),d),e)$.  The seven distinct rooted gene trees depicted in \textbf{(a)}--\textbf{(g)} all correspond to the same unrooted gene tree $T_{15}$ shown in \textbf{(h)}.  The rooted gene trees in \textbf{(c)} and \textbf{(d)} can only occur for this species tree if all coalescences occur above the root, in the population with the lightest shading.  The rooted gene trees in \textbf{(a)}, \textbf{(b)}, \textbf{(e)}, \textbf{(f)}, and \textbf{(g)} can occur with coalescent events either all above the root or with some event in the population immediately descended from the root. Only one coalescent scenario is shown for each of the rooted gene trees.}
\end{figure*}

Methods for inferring rooted species trees from multiple genes have been developed which make use of \emph{rooted} gene trees, topological or metric, which possibly differ from that of the species tree.  Most commonly, such methods assume that the incongruent gene trees (\emph{i.e.}, gene trees with topologies different from the species tree) arise because of \emph{incomplete lineage sorting}, the phenomenon that the most recent common ancestor for two gene copies is more ancient than the most recent population ancestral to the species from which the genes were sampled. Examples are shown in Fig.~\ref{F:T15}a-g, in which the lineages sampled from species $a$ and $b$ do not coalesce in the population immediately ancestral to $a$ and $b$.  Several approaches for inferring species trees in this setting have been proposed, such as minimizing deep coalesce \citep{maddison2006}, BEST \citep{liu2007}, ESP \citep{carstens2007}, STEM \citep{kubatko2009}, Maximum Tree \citep{liu2010} (also called the GLASS tree \citep{mossel2008}), and *BEAST \citep{heled2010}.  The analysis of incomplete lineage sorting requires thinking of rooted trees 
(the idea of an event being ``more ancient" requires that time have a direction), and is modeled probabilistically using coalescent theory \citep{hudson1983a,kingman1982a,nordborg2001,tajima1983,wakeley2008}.

The coalescent process was first developed to model ancestry of genes by a tree embedded within a single population, and uses exponential waiting times (going backwards in time) until two lineages coalesce.  By conceptualizing a species tree as a tree of connected populations (\emph{cf.} Fig.~\ref{F:T15}), each with its own  coalescent process, the multispecies coalescent can model probabilities of rooted gene tree topologies within a rooted species tree \citep{degnan2009,degnan2005,nei1987,pamilo1988,takahata1989,rosenberg2002}.  Although much of the work of this area has focused on one gene lineage sampled per population, extensions to computing gene tree probabilities when more than one lineage is sampled from each population has also been derived \citep{takahata1989,rosenberg2002,degnan2010}.

Under the multispecies coalescent, the species tree is a parameter, consisting of a rooted tree topology with strictly positive edge weights (branch lengths) on all interior edges.  Pendant edge weights are not specified when there is only one gene sampled per species, because it is not possible for coalescent events to occur on these edges.  Rooted gene tree topologies are treated as a discrete random variable whose distribution is parameterized by the species tree, with a state space of size $(2n-3)!! = 1 \times 3 \times \cdots \times (2n-3)$, the number of rooted, binary tree topologies \citep{felsenstein2004} for $n$ extant species (leaves). (Nonbinary gene trees are not included in the sample space since the coalescent model assigns them probability zero.)

Results on rooted triples (rooted topological trees obtained by considering subsets of three species) imply that the distribution of rooted gene tree topologies identifies the rooted species tree topology \citep{degnanEtAl2009}, in spite of the fact that the most likely $n$-taxon gene tree topology need not  have the same topology as the species tree for $n>3$ \citep{degnan2006}.  
Internal branch lengths on the species tree can also be recovered using probabilities of rooted triples from  gene trees.  
In particular, for a 3-taxon species tree in which two species $a$ and $b$ are more closely related to each other than to $c$, let $t$ denote the internal branch length.  If $p$ is the known probability that on a random rooted topological gene tree, genes sampled from species $a$ and $b$ are more closely related to each other than either is to a gene sampled from $c$,  then 
$t = -\log((3/2)(1-p))$ \citep{nei1987,wakeley2008}. Thus, for each population (edge) $e$ of the species tree, choosing two leaves whose most recent ancestral population is $e$ and one leaf descended from the immediate parental node of $e$, the length of $e$ can be determined.  We summarize these results as:
\begin{prop}\label{prop:rootedtriple}
For a species tree with $n \ge 3$ taxa, the probabilities of rooted triple gene tree topologies determine the species tree topology and internal branch lengths.
\end{prop}

Because the probability of any rooted triple is the probability that a rooted gene tree displays the triple, we have the following.
\begin{cor} \label{cor:rgt}
For a species tree with $n \ge 3$ taxa, the distribution of gene tree topologies determines the species tree topology and internal branch lengths. 
\end{cor}

\smallskip

Although previous work on modeling gene trees under the coalescent has assumed that trees are rooted, the event that a particular unrooted topological gene tree is observed can be regarded as the event that any of its rooted versions occurs at that locus \citep{heled2010}.  For $n$ species, there are $(2n-5)!! $ unrooted gene trees, and each unrooted gene tree can be realized by $2n-3$ rooted gene trees,  corresponding to choices of an edge on which to place the root.  The probability of an $n$-leaf unrooted gene tree is therefore the sum of $2n-3$ rooted gene tree probabilities, and the unrooted gene tree probabilities form a well-defined probability distribution.

In this paper, we study aspects of the distribution of unrooted topological gene trees that arises under the multispecies coalescent model
on a species tree, with the goal of understanding what one may hope to infer about the species tree.
We find that when there are only four species, with one lineage sampled from each, the most likely unrooted gene tree topology has the same unrooted topology as the species tree; however, it is impossible to recover the rooted topology  of the species tree, or all information about edge weights, from the distribution of gene trees.  When there are 5 or more species, the probability distribution on the unrooted gene tree topologies identifies the rooted species tree and all internal edge weights.  If multiple samples are taken from one of more species, then those pendant edge weights become identifiable, and the total number of taxa required for identifying the species tree can be reduced. 

In the main text, we derive these results assuming binary 
--- fully resolved --- species trees.  However, the results generalize to 
nonbinary species trees, which have internal nodes of outdegree 
greater  than or equal to 2.  Details for nonbinary cases are given 
in Appendix \ref{sec:app-nonbinary}. Implications for data analysis will 
be discussed in Section \ref{sec:Discuss}.

\medskip

We briefly indicate our approach.
Because the distribution of the $(2n-3)!!$ (rooted) or $(2n-5)!!$ (unrooted) gene trees is determined by the species tree topology and its $n-2$ internal branch lengths, gene tree distributions are highly constrained under the multispecies coalescent model.  Calculations show that many gene tree probabilities are necessarily equal, or satisfy more elaborate polynomial constraints.  Polynomials in gene tree probabilities which evaluate  to 0 for any set of branch lengths on a particular species tree topology are called \emph{invariants} of the gene tree distribution for that species tree topology.
A trivial example, valid for any species tree, is that the sum of all gene tree probabilities minus 1 equals 0.  Many other invariants express  ties in gene tree probabilities.  For example, consider the rooted species tree $((a,b),c)$, where $t$ is the length of the internal branch.  Suppose gene $A$ is sampled from species $a$, $B$ from $b$, and $C$ from $c$.  Then the rooted gene tree $((A,B),C)$ has probability $p_1 = 1-(2/3)\exp(-t)$ under the coalescent, while the two alternative gene trees, $((A,C),B)$ and $((B,C),A)$, have probability $p_2 = p_3 = (1/3)\exp(-t)$ \citep{nei1987}.  Thus  a rooted gene tree invariant for this species tree is 
\begin{equation}\label{E:invariant3taxa}
p_2 - p_3 =0.
\end{equation}
We emphasize that this invariant holds for all values of the branch length $t$.  The species tree also implies certain inequalities in the gene tree distribution; for example, for any branch length $t>0$, $p_1 > p_2$.  Because of such inequalities, the invariant in equation (\ref{E:invariant3taxa}) holds on a gene tree distribution if, and only if, the species tree has topology $((a,b),c)$.

Different species tree topologies imply different sets of invariants and inequalities for their gene tree distributions, for both rooted and unrooted gene trees.  We note that previous work on invariants for statistical models in phylogenetics \citep{allman2003,cavender1987,lake1987} has focused on polynomial constraints for site pattern probabilities; that is, probabilities that leaves of a gene tree display various states (e.g., one of four states for DNA nucleotides) under models of character change, given the topology and branch lengths of the gene tree.   These approaches have been particularly useful in determining identifiability of (gene) trees given sequence data under different models of mutation \citep{AHRfilter,APRS2tree,ARidtree,ARGMI,ARcov}.

In this paper, our methods focus on understanding linear invariants  and inequalities for distributions of unrooted gene tree topologies under the multispecies coalescent model.  Here gene trees are branching patterns representing ancestry and descent for genetic lineages, and are independent of mutations that may have arisen on these lineages.  This is therefore a novel application of invariants in phylogenetics.

\section{Notation}

\begin{figure*}[h]\label{F:modeltrees}
\begin{center}
\includegraphics[width=.3\textwidth]{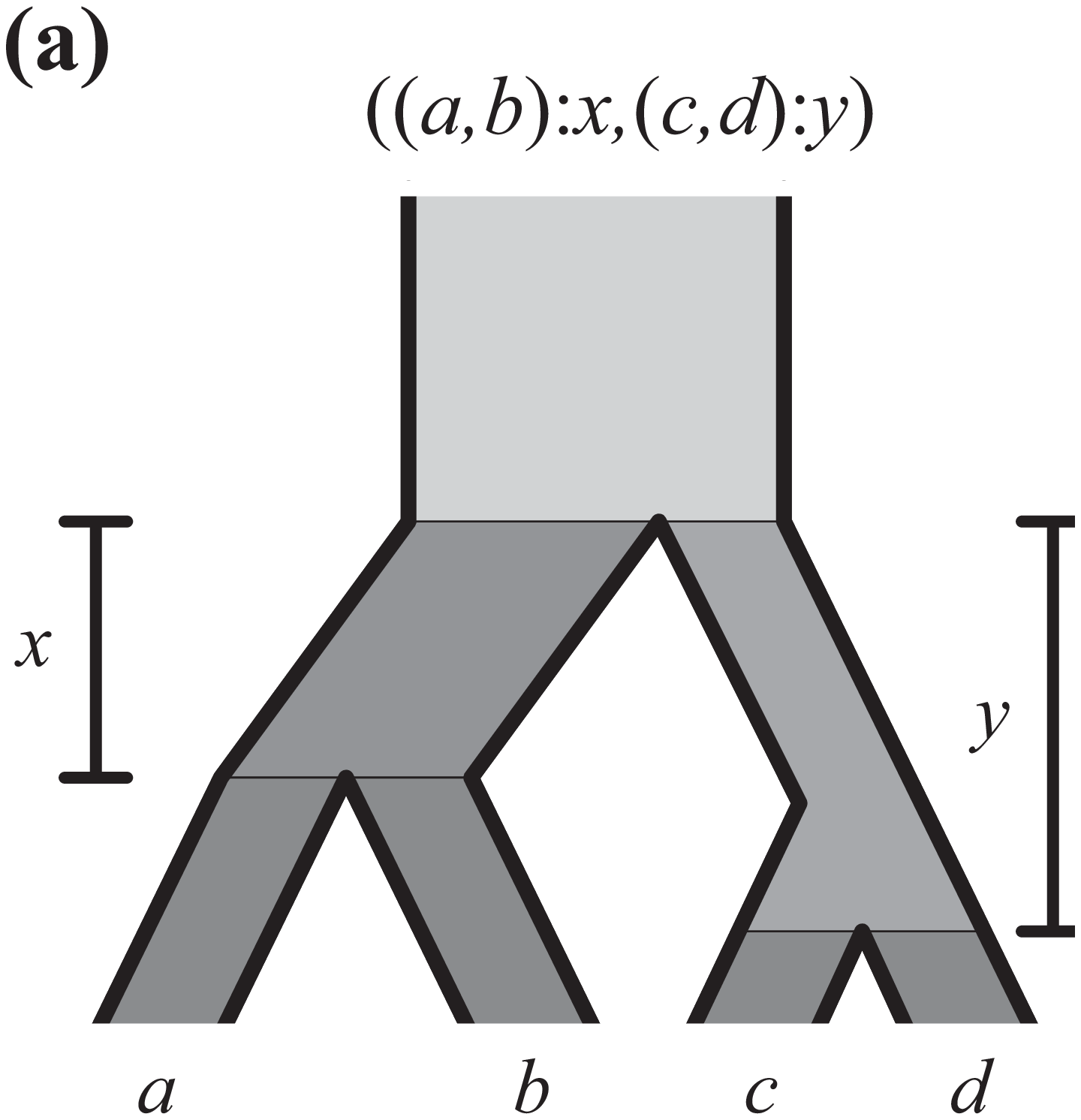}
\includegraphics[width=.3\textwidth]{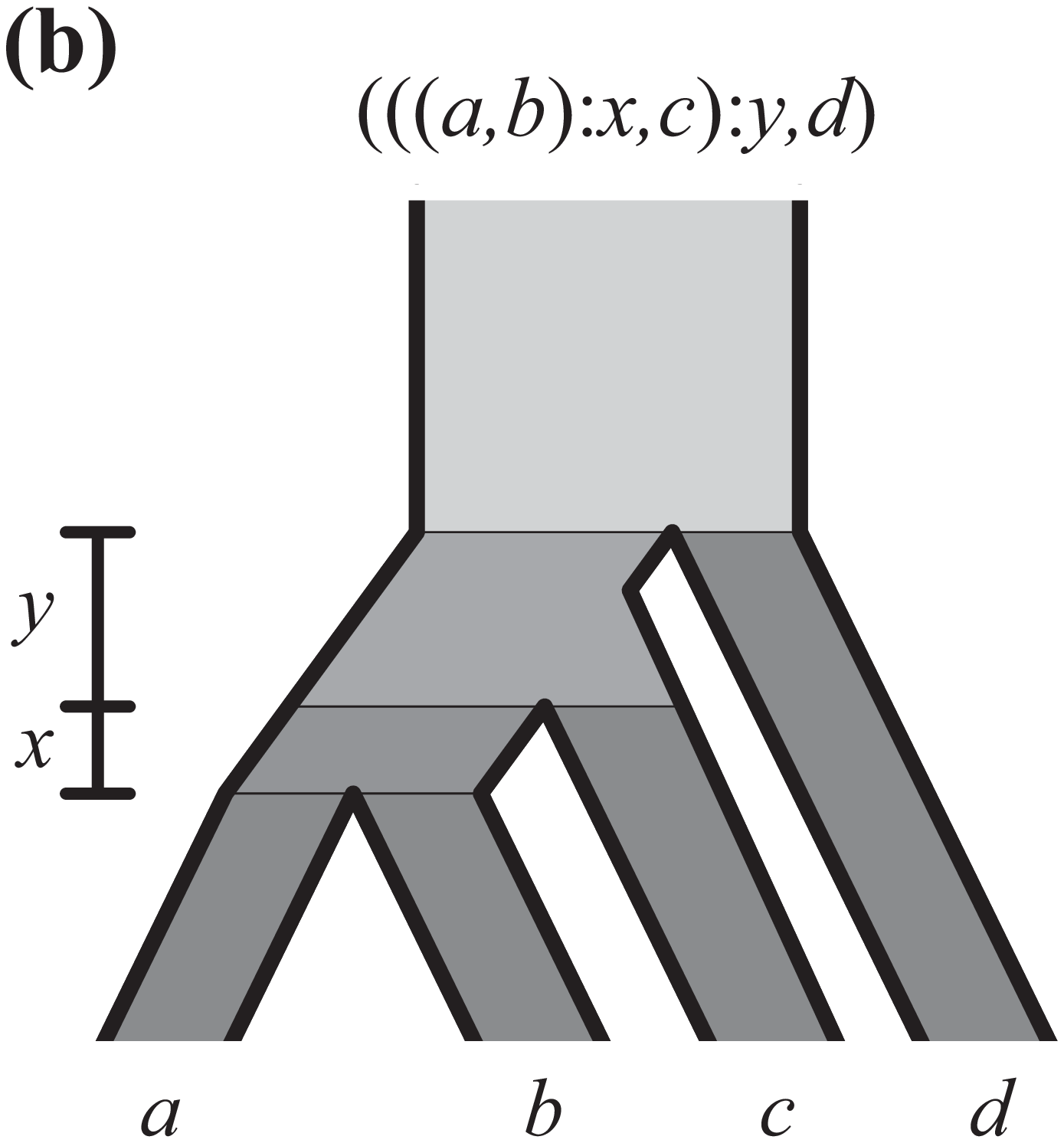}\\
\vskip 4mm
\includegraphics[width=.3\textwidth]{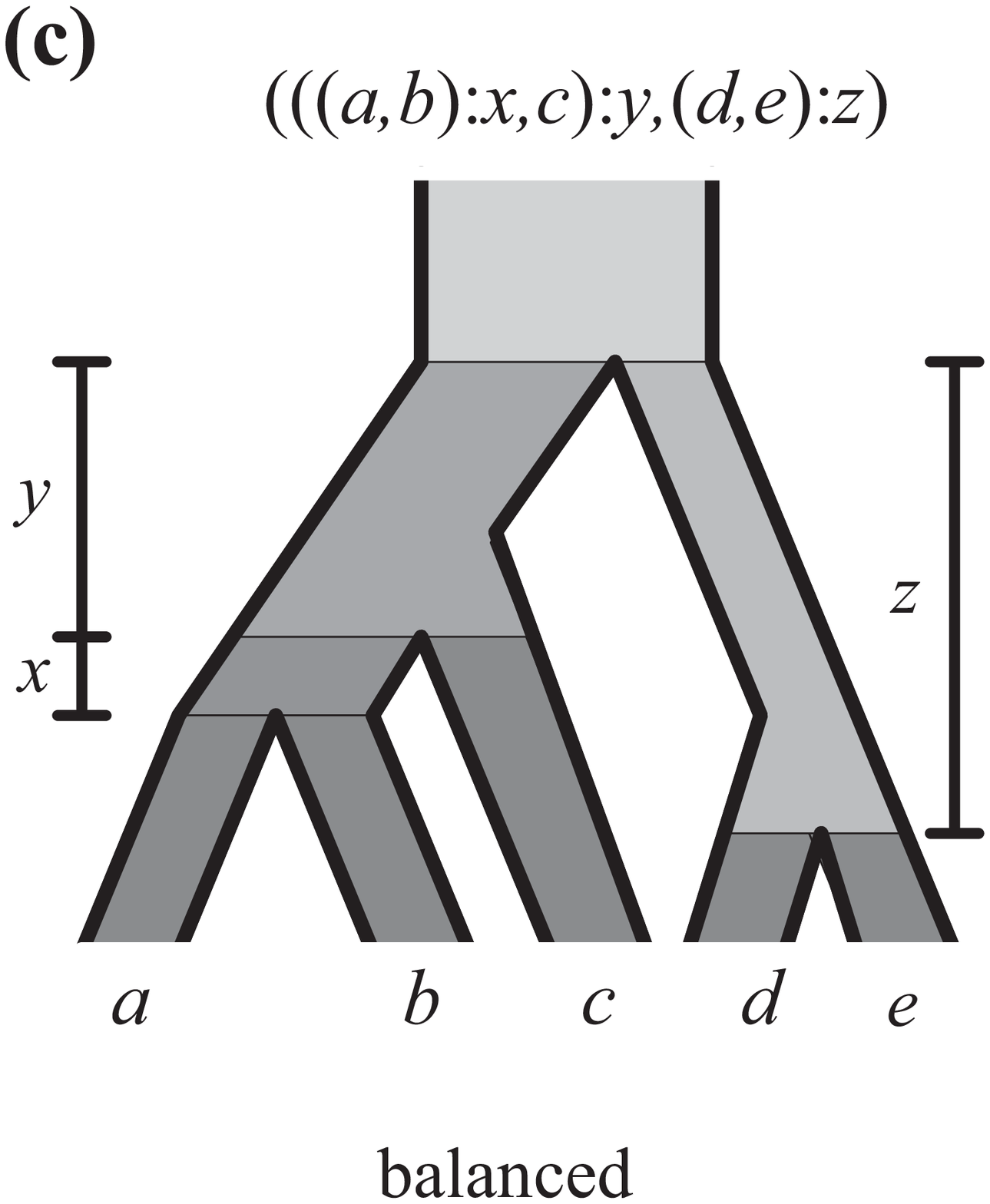}
\hskip 2mm
\includegraphics[width=.3\textwidth]{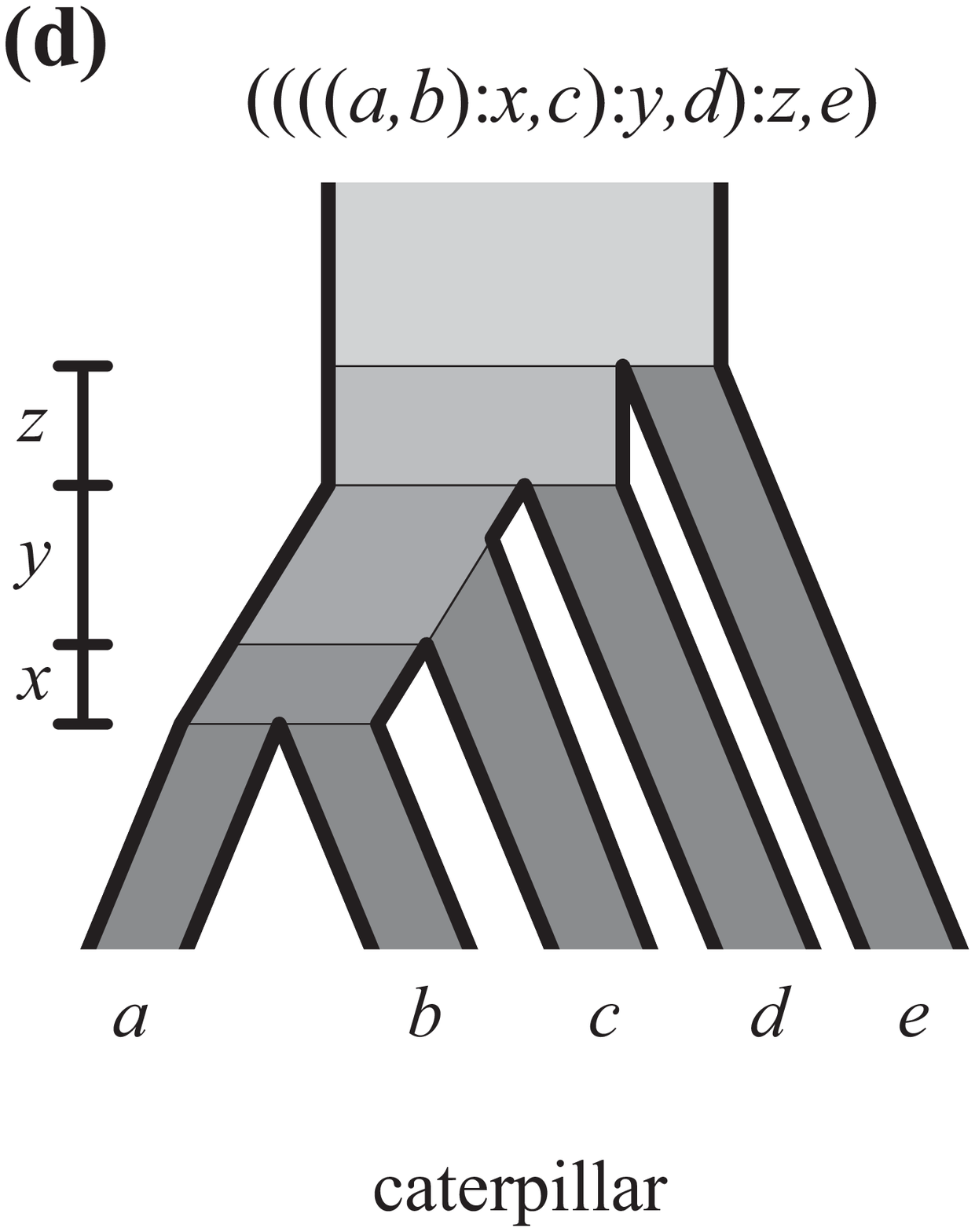}
\hskip 2mm
\includegraphics[width=.3\textwidth]{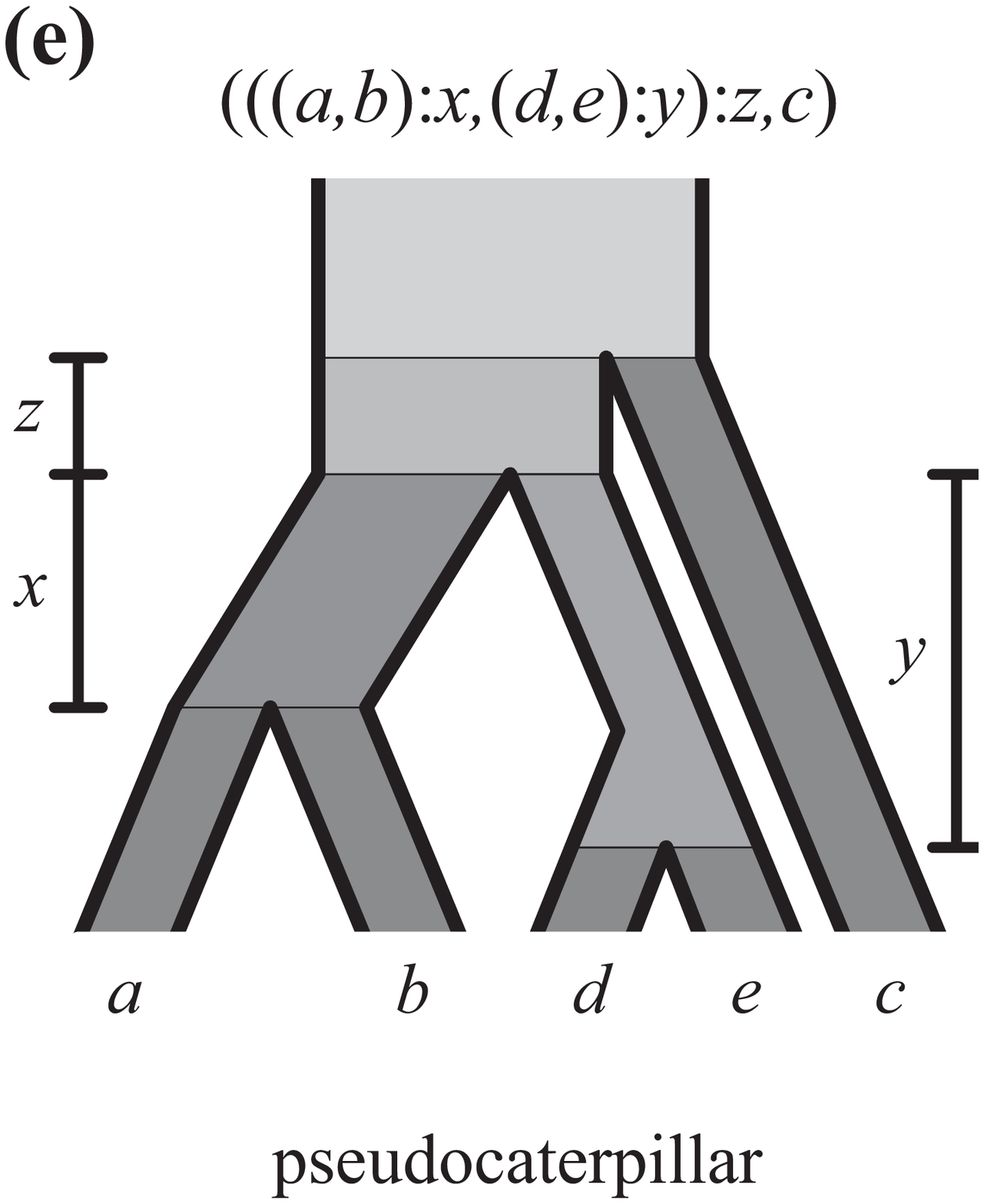}
\end{center}
\caption{Model species trees with branch lengths used to determine  probabilities of unrooted gene trees in this paper.  The two 4-taxon species trees in \textbf{(a)} and \textbf{(b)} each have the same unrooted topology, namely a tree with the $ab|cd$ split.  The three 5-taxon species trees in \textbf{(c)}--\textbf{(e)} also share one unrooted topology, the topology with the splits $ab|cde$ and $abc|de$. }
\end{figure*}

Let $X$ denote a set of $|X|=n$ taxa, and let
$\psi^+$  denote a rooted, binary,  topological species tree whose $n$ leaves are labeled by the elements of $X$. If $\psi^+$ is further endowed with a collection $\lambda^+$ of strictly positive edge lengths for the $n-2$ internal edges, then
$\sigma^+=(\psi^+,\lambda^+)$ denotes a rooted, binary, leaf-labeled, metric species tree on $X$.
Note that edge lengths in the species tree do not represent evolutionary time directly, but are in 
{\it coalescent units}, that is units of $\tau/N_e$, where $\tau$ is the number of generations and 
$N_e$ is the effective population size, the effective number of gene copies in a population \citep{degnan2009}.
As pendant edge lengths do not affect the probability of observing any topological gene tree, rooted or unrooted, under the multispecies coalescent model with one individual  sampled from each taxon, they are not specified in $\lambda^+$. 
To specify a particular species tree $\sigma^+$, we use a modified Newick notation which omits pendant edge lengths. For instance, a particular 4-taxon balanced metric species tree is $\sigma^+=((a,b)\tc 0.1,(c,d)\tc0.05)$.  Rooted 4- and 5-taxon species trees with branch lengths which will be used later in this paper are depicted in Fig.~\ref{F:modeltrees}.  We refer to the 5-taxon tree shapes as \emph{balanced}, \emph{caterpillar}, and following Rosenberg\nocite{rosenberg2007}, \emph{pseudocaterpillar}.

Replacing `$+$' with `$-$' denotes suppressing the root, so that
$\psi^-$ is the unrooted binary topological species tree, $\lambda^-$ the induced collection of $n-3$ internal edge lengths on $\psi^-$, and $\sigma^-=(\psi^-,\lambda^-)$ is the unrooted metric species tree.  An unrooted topology can be specified by its nontrivial \emph{splits} --- the partitions of the taxa induced by removing an internal edge of the unrooted tree.  For example, $T_{15}$ in Fig.~\ref{F:T15}h has splits $BE|ACD$ and $ABE|CD$.   A set of all taxa descended from a node in a rooted tree forms a \emph{clade}, the rooted analog of a split.  For example, the rooted gene tree in Fig.~\ref{F:T15}a has 2-clades $\{B,E\}$ and $\{C,D\}$ and the 3-clade $\{A,B,E\}$. 

For any set of taxa $S\subseteq X$, we let $\mathcal{T}_S$ denote the collection of all unrooted, binary, leaf-labeled topological gene trees for the taxa $S$. We use the convention that while lower-case letters denote taxa on a species tree, the corresponding upper-case letters are used
as leaf labels on a gene tree; Thus $A$ denotes a gene from taxon $a$, etc. 
For example, if $X=\{a,b,c,d\}$, then 

$$\mathcal T_{X}=\{ AB|CD,\,AC|BD,\,AD|BC\}.$$

Given any sort of tree (species/gene, rooted/unrooted, topological/metric) on $X$, appending `$(S)$' denotes the induced tree on the taxa $S\subseteq X$.
By `induced tree' here we mean the tree obtained by taking the minimal subtree with leaves in $S$ and then suppressing  all non-root nodes of degree 2.
Instances of this notation include $\sigma^+(S)$, $\sigma^-(S)$, $\psi^+(S)$, $\psi^-(S)$, and $T(S)$.

\section{The multispecies coalescent model}\label{sec:coalmodel}

Several papers have given examples of applying the coalescent process to  multiple species or populations  to derive examples of probabilities of rooted gene tree topologies given species trees \citep{nei1987,pamilo1988,rosenberg2002} with the general case (for any $n$-taxon, rooted, binary species tree) given in \citep{degnan2005}.  We present  the model here with only one individual sampled per taxon, as that will be sufficient for our analysis.

Under the multispecies coalescent model, waiting times (going backwards in time) until coalescent events (nodes in a rooted gene tree) are exponential random variables. The rate for these variables is  $\binom{i}{2}$, with $i$ the number of lineages ``entering" a population, i.e., a branch on the species tree. 
Gene tree probabilities can be computed by enumerating all possible specifications of branches in which each coalescent event occurs, and computing the probability of these events in each branch, treating each branch as a separate population.   In particular, the probability that $i$ lineages coalesce into $j$ lineages within time $t$ is represented by the function $g_{ij}(t)$ \citep{tavare1984}, which is a linear combination of exponential functions:
\begin{equation}
g_{ij}(t) = \sum_{k=j}^i \exp \left ( {-\binom{k}{2}t}\right )\frac{(2k-1)(-1)^{k-j}}{j!(k-j)!(j+k-1)}\prod_{m=0}^{k-1}\frac{(j+m)(i-m)}{i+m}, \;\;\; 1 \le j \le i.
\end{equation}
Here $t>0$ is time measured in coalescent units.  The functions $g_{ij}$ have the property that for any $i>1$ and any $t>0$, $g_{ij}(t)$, $j=1, \dots, i$, is a discrete probability distribution, that for any $i>1$, $\lim_{t \rightarrow \infty} g_{i1}(t) = 1$, and that $\lim_{t \rightarrow 0} g_{ii}(t) = 1$.  These last two properties express the ideas that given enough time, all lineages eventually coalesce (there is only one lineage remaining in a population) and that over very short time intervals, it is very likely that no coalescent events occur.  Finally, note that  $g_{ii}(t) = \exp\left(-i(i-1)t/2\right )$.

As an example of using this function to determine rooted gene tree probabilities, consider the rooted caterpillar species tree $((((a,b)\tc x,c)\tc y,d)\tc z,e)$
of Fig.~\ref{F:modeltrees}d, and the rooted gene tree $((((B,E),A),C),D)$. Since this gene tree requires a specific ordering of coalescences, and the first of these can only occur in the population above the root of the species tree, the only scenario to consider is that shown in Fig.~\ref{F:T15}c.  In the population ancestral to species $a$ and $b$, there are two lineages which must fail to coalesce in time $x$, and this event has probability $g_{22}(x) = \exp(-x)$.  Similarly, the events in the populations with durations $y$ and $z$ have probabilities $\exp(-3y)$ and $\exp(-6z)$, respectively, because no lineages coalesce in those intervals.  For the population ancestral to the root, all lineages eventually coalesce, and the probability for events in this population is the probability of observing the particular sequence of coalescence events, which is $\left (\binom{5}{2}\binom{4}{2}\binom{3}{2}\binom{2}{2} \right )^{-1} = 1/180$.  The probability of the rooted gene tree given the species tree is therefore $\exp(-x-3y-6z)/180$.  It is often convenient to work with transformed branch lengths, where if a branch has length $x$, we set $X = \exp(-x)$.  Using this notation, the rooted gene tree has probability $XY^3Z^6/180$.

As another example, consider the gene tree $(((B,E),A),(C,D))$
given the same species tree,
$((((a,b)\tc x,c)\tc y,d)\tc z,e)$. 
For this rooted gene tree to be realized, either $C$ and $D$ coalesce as depicted in Fig.~\ref{F:T15}a, in the population immediately below the root (which we call the ``near the root'' population), or $C$ and $D$ coalesce above the root.  Regardless, all other coalescent events must occur in the population above the root.  We therefore divide the calculation of the rooted gene tree topology into these two cases. If all coalescent events occur above the root, the rooted gene tree probability is calculated as in the preceding paragraph, except that there are three possible orders in which the coalescent events could occur to realize the rooted gene tree, and the probability for this case is thus $XY^3Z^6/60$.
In the case where $C$ and $D$ coalesce ``near the root,"  there are no coalescent events in the populations with lengths $x$ and $y$, thus contributing a factor of $\exp(-x-3y)$ to the probability.  The probability for events near the root is $ \binom{4}{2}^{-1}g_{43}(z)$, where the coefficient is the probability that of the four lineages entering the population, the two that coalesce are $C$ and $D$.  Because there are four lineages entering the population above the root of the species tree, the  one sequence of coalescent events that results in the gene tree topology has probability $\left (\binom{4}{2}\binom{3}{2}\binom{2}{2}\right )^{-1} = 1/18$.  The total probability of the rooted gene tree topology $(((B,E),A),(C,D))$ given the species tree $(((a,b)\tc x,c)\tc y,d)\tc z,e)$ is therefore
\begin{align*}
g_{22}(x)g_{33}(y)\frac{1}{\binom{4}{2}}g_{43}(z)\frac{1}{\binom{4}{2}\binom{3}{2}\binom{2}{2}} + g_{22}(x)g_{33}(y)&g_{44}(z)\frac{3}{\binom{5}{2}\binom{4}{2}\binom{3}{2}\binom{2}{2}}\\
 = &XY^3\frac{1}{6}(2Z^3-2Z^6) \frac 1{18}+ \frac{1}{60}XY^3Z^6\\
 = &\frac{1}{54}XY^3Z^3 - \frac{1}{540}XY^3Z^6.
 \end{align*}

 Probabilities of the other rooted gene trees in Fig.~1 can be worked out similarly by considering a small number of cases for each tree.  
 Methods for enumerating all possible cases have been developed using the concept of \emph{coalescent history}, a list of populations in which the coalescent events occur  \citep{degnan2005}.  Each coalescent history $\mathbf h$ has a probability of the form
 \begin{equation}\label{E:phist}
 c(\mathbf{h}) \prod_{b=1}^{n-2} g_{i(\mathbf{h},b),j(\mathbf{h},b)}(x_b)
 \end{equation}
where $x_b$ is the length of internal edge $b$ of the species tree and $c(\mathbf{h})$ is a constant that depends on the coalescent history $\mathbf{h}$ and the topologies of the gene and species trees, but does not depend on the branch lengths $x_b$. This expression is a linear combination of products of terms $\exp[-k(k-1)x_b/2]$, $k=2,\dots,n-1$, so using the transformations $X_b = \exp(-x_b)$,  probabilities of coalescent histories can thus be written as polynomials in the transformed branch lengths of the species tree.  Because gene tree probabilities are sums of probabilities of coalescent histories, gene tree probabilities can also be written as polynomials in the transformed branch lengths.  

Finally, unrooted gene tree probabilities, which are sums of rooted gene tree probabilities, can also be expressed as polynomials in the transformed branch lengths.
We thus can derive polynomial expressions for the probabilities of unrooted gene trees given a species tree.

\section{Results}

The unrooted topological gene tree distribution under the multispecies coalescent model on species tree $\sigma^+$, with one lineage sampled per species, will be denoted by $ \mathbb P=\mathbb P_{\sigma^+}$, so that
$\mathbb P(T)$ denotes the probability of observing gene tree $T\in \mathcal T_{X}$.  

For ease of exposition, we assume throughout this section that the species tree $\sigma^+$ is binary.
See Section \ref{sec:nonbinary} for the polytomous case.

\subsection{4-taxon trees}\label{sec:4taxa}
We first consider the case of four taxa, and so let $X=\{a,b,c,d\}$.  
Using non-trivial splits as indices, the set of  gene trees is $$\mathcal T_X=\{T_{AB|CD}, T_{AC|BD}, T_{AD|BC}\}.$$

With four taxa, there are only two shapes for species trees: the balanced tree, with two clades of size 2 (Fig.~\ref{F:modeltrees}a); and the rooted caterpillar tree with a 2-clade nested inside a 3-clade (Fig.~\ref{F:modeltrees}b). Of the 15 possibilities for $\psi^+$, there are three labeled balanced tree topologies, and 12 labeled caterpillar topologies. It is only necessary to compute gene tree probabilities for a single labeling of the leaves of each species tree shape, since permuting labels immediately gives the distribution for other choices.

For a balanced tree $\sigma^+=(((a,b)\tc x,(c,d)\tc y)$ shown in Fig.~\ref{F:modeltrees}a, one computes, as described in the previous section, that the gene tree distribution is given by

\begin{align*}
\mathbb P_{\sigma^+} (T_{AB|CD})&=1-\frac 23 e^{-(x+y)},\\
\mathbb P_{\sigma^+} (T_{AC|BD})=\mathbb P_{\sigma^+} &(T_{AD|BC})=\frac 13 e^{-(x+y)}.
\end{align*}

For a rooted caterpillar species tree $\sigma^+=(((a,b)\tc x,c)\tc y,d)$ shown in Fig.~\ref{F:modeltrees}b, one finds

\begin{align*}
\mathbb P_{\sigma^+} (T_{AB|CD})&=1-\frac 23 e^{-x},\\
\mathbb P_{\sigma^+} (T_{AC|BD})=\mathbb P_{\sigma^+} &(T_{AD|BC})=\frac 13 e^{-x}.
\end{align*}

Thus for any 4-taxon species tree, from the distribution $\mathbb P_{\sigma^+}$ 
 one can identify the unrooted species tree topology $\psi^-$ as that of the most probable unrooted gene tree $T$. The
one internal edge length on $\psi^-$ (\emph{i.e.}, $x+y$ in the balanced case, $x$ for the caterpillar) can be recovered as $-\log\left(\frac 32\left (1-\mathbb P(T)\right)\right)$. Thus $\sigma^-=(\psi^-,\lambda^-)$ is identifiable.

Furthermore $\sigma^+$ is  not identifiable since the above calculations show that for any $x>0$, $y_i>0$, and $x>z>0$ the following rooted species trees
produce exactly the same unrooted gene tree distribution:

\begin{align*}
&((a,b)\tc x,c)\tc y_1,d),\\
&((a,b)\tc x,d)\tc y_2,c),\\
&((c,d)\tc x,a)\tc y_3,b),\\
&((c,d)\tc x,b)\tc y_4,a),\\
&((a,b)\tc z,(c,d)\tc x-z).
\end{align*}

We summarize this by:

\begin{prop} \label{prop:4taxa} For $|X|=4$ taxa, $\sigma^-$ is identifiable from $\mathbb P_{\sigma^+}$, but $\sigma^+$ is not.
\end{prop}

We note that if the unrooted gene trees are ultrametric with known branch lengths, then their rooted topologies are known by midpoint rooting \citep{kim1993}, and thus $\sigma^+$ is identifiable from unrooted ultrametric 4-taxon gene trees.

\subsection{ Linear invariants and inequalities for unrooted gene tree probabilities for 5-taxon species trees}\label{sec:lininv} 

To establish identifiability of all parameters when there are at least 5 taxa, we will argue from the 5-taxon case. In this base case
we will use an understanding of linear relationships --- both equalities and inequalities --- that hold between gene tree probabilities. The relationships that hold for a particular gene tree distribution reflect the species tree on which it arose.

In this section, we determine all linear equations in gene tree probabilities for each of the three  shapes of 5-leaf species trees. Following phylogenetic terminology, these are the \emph{linear invariants} of the gene tree distribution. We emphasize that these invariants depend only on the rooted topology, $\psi^+$, of the species tree, and not on the branch lengths $\lambda^+$. Although some of these invariants arise from symmetries of the species tree, others are less obvious. Nonetheless, we give simple arguments for all, and show that there are no others. In addition, we provide all pairwise inequalities of the form $u_i > u_j$ for the three model species trees in Figs.~\ref{F:modeltrees}c--e.

With $X=\{a,b,c,d,e\}$, there are 15 unrooted gene trees in $\mathcal T_X$, which we enumerate in Table \ref{T:5taxaprobs} of Appendix \ref{app:tables}.  Probabilities for each of the 15 unrooted gene trees are obtained by summing probabilities of seven of the 105 rooted 5-taxon gene trees, as shown in Tables \ref{T:5taxarooted} and \ref{T:5taxaprobs} of Appendix \ref{app:tables}.
In Appendix \ref{app:gtdist} formulas for the unrooted gene tree distribution are given for one choice of a leaf-labeling of each of the three possible rooted species tree shapes. Noticing that many of the gene tree probabilities are equal, one might hope that which ones are equal would be useful in identifying the species tree from the distribution.

For each species tree one can computationally, but entirely rigorously, determine a basis for the vector space of all linear invariants. We report such a basis for each of the species tree shapes below, in Tables \ref{T:explanation_balanced}-\ref{T:explanation_pseudo}. Only for one of the tree shapes is an additional invariant that is not immediately noticeable produced by this calculation.  While our computations were performed using the algebra software Singular \citep{GPS09}, many other packages would work as well, or one could do the calculations without machine aid. 

In the tables and discussion below, we omit mention of the trivial invariant, $$\sum_{i=1}^{15} \mathbb P_{\sigma^+}(T_i)=1,$$
which holds for any choice of $\sigma^+$. We instead only give a basis for the homogeneous linear invariants.

We use the following observation.

\begin{lem}\label{lem:aboveroot}
If all coalescent events occur above the root (temporally before the MRCA of all species) of a 5-taxon species tree, then all 15 of the unrooted topological gene trees are
equally likely.
\end{lem}
\begin{proof} If all coalescent events occur above the root, then regardless of the species tree, we are considering five labeled lineages
entering the ancestral population, and then coalescing. 
Because all unrooted gene trees have the same unlabeled shape, all coalescent histories leading to one gene tree correspond to equally likely coalescent histories  producing another, by simply relabeling lineages.  \qed
\end{proof}
Note that the claim of this lemma is special to five taxa. For six taxa, with two different unrooted gene tree shapes possible, the analogous statement is not true.

\subsubsection{Balanced species tree}
Suppose $\psi^+=(((a,b),c),(d,e))$, as depicted in Fig.~\ref{F:modeltrees}c. Because $\sigma^+$ is invariant under interchanging taxa $a$ and $b$, any two gene trees that differ by 
transposing leaves $A$ and $B$
must have the same probability. Similarly, interchanging $D$ and $E$ on a gene tree cannot change its probability.
We refer to the first permutation of labels using cycle notation as $(ab)$, and the second as $(de)$. More formally, assuming generic values for $\lambda^+$, the symmetry group of $\sigma^+$   is the 4-element group generated by the transpositions $(ab)$ and $(de)$, and the gene tree probability distribution must be invariant under the action of this group on gene trees. These symmetries thus give `explanations' for many invariants holding.

A different explanation for some invariants is that some unrooted gene trees can only be realized if all coalescent events occur above (more anciently than) the root of the species tree.  For example, any realization of the gene tree $T_{15}$ with splits $BE|ACD$ and $ABE|CD$ (Fig.~\ref{F:T15}h) requires that the first (most recent) coalescent event either joins lineages $B$ and $E$, or  joins $C$ and $D$. Because both of these events can only occur above the root, all events must take place above the root.
Another such gene tree is $T_{11}$, with splits $AE|BCD$ and $ACE|BD$.   
Thus by Lemma \ref{lem:aboveroot}
the unrooted gene trees $T_{11}$ and $T_{15}$
must have the same probability, even though they do not differ by a symmetry as described in the last paragraph. We refer to this reasoning as the ``above the root'' argument.

Some invariants can be explained in several ways. For example, the same invariant might be explained by two different symmetries or by both a symmetry and an above-the-root argument. In Table \ref{T:explanation_balanced}, we list a basis for homogeneous linear invariants, and give only one explanation for each. 
Here $u_i=\mathbb P(T_i)$.

\begin{table}[!h]
\caption{Invariants for the rooted species tree $\psi^+=(((a,b),c),(d,e))$}
\label{T:explanation_balanced}
\begin{center}
\begin{tabular}{cc}
Invariant& Explanation\\
\hline
$u_{14}-u_{15}=0$    &   $(de)$\\
$u_{11}-u_{15}=0$   & above root\\
$u_{10}-u_{15}=0$    &   $(ab)$\\
$u_9-u_{12}=0$      & $(de)$\\
$u_8-u_{15}=0$      &  above root\\
$u_7-u_{15}=0$      & $(ab)(de)$\\
$u_6-u_{12}=0$      & $(ab)(de)$\\
$u_5-u_{12}=0$     & $(ab)$\\
$u_4-u_{13}=0$      & $(ab)$\\
$u_2-u_3=0$      & $(de)$\\
\hline
\end{tabular}
\end{center}
\end{table}

These equalities give the following equivalence classes of unrooted gene trees according to their probabilities:
$$\{T_{1}\}, \{T_{2}, T_{3}\},\{T_4,T_{13}\}, \{T_{5},T_{6},T_9,T_{12}\}, \{T_{7},T_{8},T_{10},T_{11}, T_{14},T_{15} \}.$$

For any branch lengths on this species tree, we also observe the inequalities 
\begin{equation}\label{E:inequalities1}
u_1>u_2,u_4>u_5>u_7.
\end{equation}

These inequalities were found by first expressing the probability of each $T_i$ as a sum of positive terms corresponding to coalescent histories, such as expression \eqref{E:phist}, and then, by comparing coefficients in these sums, determining instances in which $u_i>u_j$ must hold.
Intuitively, this means that any realization of $T_j$ corresponds to a realization of $T_i$, but that there are additional ways that $T_i$ can be realized.

The inequalities in (\ref{E:inequalities1}) can all be checked by elementary arguments using the explicit formulas of Appendix \ref{app:gtdist}. For instance, since $X,Y,Z\in(0,1)$,
\begin{align*}
u_1-u_2 &= 1 - \frac{2}{3}X - YZ + \frac{1}{2}XYZ + \frac{1}{6}XY^3Z= 1-YZ-\frac 16X(  4 - 3YZ- Y^3Z)\\
                 &>  1-YZ-\frac 16(  4 - 3YZ- Y^3Z)=  \frac13 - \frac{1}{6}YZ(3 - Y^2)\\
                 &>\frac 13-\frac 16Y(3-Y^2)>0.
\end{align*}

In particular, there is always a 6-element equivalence class of trees which has the strictly smallest probability associated with it, and a 4-element class which has the next smallest probability associated to it. While the class associated to the largest probability is always a singleton, these inequalities do allow for the remaining two classes of size 2 to degenerate to a single class of size 4.  

Numerical examples can be used to show that there are no inequalities of the form $u_i > u_j$ that hold for all branch lengths $X$, $Y$, and $Z$ that are not listed in (\ref{E:inequalities1}).

\subsubsection{Caterpillar species tree}\label{sec:lininvcat}
Suppose $\psi^+=((((a,b),c),d),e)$, as depicted in Fig.~\ref{F:modeltrees}d. Then the symmetry group of the tree is generated by $(ab)$, and has only two elements.

Although no unrooted gene trees require that all coalescent events occur above the root of this species tree, there are gene trees that require that all events be either above the root or ``near the root'' in the following sense. Consider the gene tree $T_{15}$ with splits $BE|ACD$ and $ABE|CD$ (Fig.~\ref{F:T15}h). This gene tree can be realized either by all events occurring above the root (in which case either the $BE$ coalescence or the $CD$ coalescence could be first), or by 
1, 2, or 3 events occurring in a specific order in the near-the-root population which is ancestral to
species $a,b,c$, and $d$ but not to $e$, with all further events above the root.  For example, if there are two coalescent events in this population, then the gene tree must
have $((CD)A)$ as a subtree (Fig.~\ref{F:T15}b,e,f), and $C$ and $D$ must coalesce most recently followed by the coalescence of $A$.  In case  1, 2, or 3 events do occur below the root, these must be in the specific order 1) $CD$ coalesce, 2) $ACD$ coalesce, 3) $ABCD$ coalesce. Another gene tree which leads to a similar 
analysis of how coalescent events must occur for the gene tree to be realized is $T_{14}$, with splits $BD|ACE$, $ABD|CE$. Consequently, $T_{14}$ has the same probability as $T_{15}$, even though these two gene trees do not differ by a symmetry.
Similar arguments apply to trees $T_7$, $T_8$, $T_{10}$, and $T_{11}$. The near-the-root argument and symmetry between $a$ and $b$ explain all linear invariants but the last in Table \ref{T:explanation_caterpillar}.

\begin{table}[!h]
\begin{center}
\caption{Invariants for the rooted species tree $\psi^+=((((a,b),c),d),e)$}
\label{T:explanation_caterpillar}
\begin{tabular}{cc}
Invariant& Explanation\\
\hline
$u_{14}-u_{15}=0$& near root\\
$u_{11}-u_{15}=0$& near root\\
$u_{10}-u_{15}=0$&$(ab)$\\
$u_8-u_{15}=0$& near root\\
$u_7-u_{15}=0$& near root\\
$u_6-u_9=0$&$(ab)$\\
$u_5-u_{12}=0$&$(ab)$\\
$u_4-u_{13}=0$&$(ab)$\\
$u_2-u_3+u_9-u_{12}=0$& marginalization\\
\hline
\end{tabular}
\end{center}
\end{table}

To explain the last invariant in Table \ref{T:explanation_caterpillar}, we provide a marginalization argument. We use the fact that for 4-taxon trees the two unrooted gene trees that are inconsistent with the species tree are equiprobable. Thus, marginalizing over $a$ to trees on $\{b,c,d,e\},$ we have that
$\mathbb P(T_{BD|CE})=\mathbb P(T_{BE|CD})$.  Hence,
\begin{equation*}
u_2+u_6+u_7+u_{11}+u_{14}=u_3+u_5+u_8+u_{10}+u_{15}.
\end{equation*}
Because the last 3 terms on each side are equal to $u_{15}$, we may cancel those.  Replacing $u_6$ with $u_9$, and $u_5$ with $u_{12}$, then gives the last invariant in the table.

Table \ref{T:explanation_caterpillar} yields the following equivalence classes of gene trees according to their probabilities:
$$\{T_{1}\},\{T_{2}\},\{T_{3}\},\{T_{4},T_{13}\},\{T_{5},T_{12}\},\{T_{6},T_{9}\},\{T_{7},T_{8},T_{10},T_{11},T_{14},T_{15}\}.$$
We also observe that the inequalities
\begin{align}\label{E:inequalities2}
u_1>u_2,&u_4>u_5>u_7,\notag\\
u_3>u_2,&u_6>u_5>u_7
\end{align}
hold for all branch lengths on this species tree, and that there are no other inequalties of the form $u_i > u_j$ that hold for all branch lengths, by arguments similar to those for the balanced tree.

\subsubsection{Pseudocaterpillar species tree}
Suppose $\psi^+=(((a,b),(d,e)),c)$, as depicted in Fig.~\ref{F:modeltrees}e. Then the symmetry group of the tree $\sigma^+$ is generated by $(ab)$ and $(de)$, and has four elements. (Note that interchanging the two cherries, for instance by $(ad)(be)$, is a symmetry of $\psi^+$, but  is \emph{not} a symmetry of $\sigma^+$ for generic edge lengths.)

While no unrooted gene trees require that all coalescent events occur above the root of this species tree, some unrooted gene trees require that all events be either near the root or above the root. The gene tree $T_{15}$, with splits $BE|ACD$ and $ABE|CD$, can be realized either by all events occurring above the root (in which case either the $BE$ coalescence or the $CD$ coalescence could be first), or by  1, 2, or 3 events occurring in a specific order in the population ancestral to 
species $a,b,c$, and $d$ but not to $e$, with all further events occurring above the root.  In case  1, 2, or 3 events do occur below the root, these must be in the specific order 1) $BE$ coalesce, 2) $ABE$ coalesce, 3) $ABDE$ coalesce. Another gene tree which leads to a similar analysis of how coalescent events must occur for the gene tree to be realized is $T_{12}$, with splits $AE|BCD$, $ADE|BC$. Thus $T_{12}$ and $T_{15}$ are equiprobable, even though they do not differ by a symmetry.

A basis for homogeneous linear invariants of unrooted gene tree probabilities, along with explanations for each is given in Table \ref{T:explanation_pseudo}.

\begin{table}[!h]
\begin{center}
\caption{Invariants for the rooted species tree $\psi^+=(((a,b),(d,e)),c)$}
\label{T:explanation_pseudo}
\begin{tabular}{cc}
Invariant& Explanation\\
\hline
$u_{14}-u_{15}=0$&$(de)$\\
$u_{12}-u_{15}=0$& near root\\
$u_{10}-u_{15}=0$&$(ab)$\\
$u_{9}-u_{15}=0$&near root\\
$u_{8}-u_{11}=0$& $(ab)$\\
$u_7-u_{15}=0$&$(ab)(de)$\\
$u_6-u_{15}=0$&near root\\
$u_5-u_{15}=0$&near root\\
$u_4-u_{13}=0$&$(ab)$\\
$u_2-u_3=0$&$(de)$\\
\hline
\end{tabular}
\end{center}
\end{table}

We thus obtain the following equivalence classes of unrooted gene trees according to their probabilities:
$$\{T_{1}\},\{T_{2}, T_{3}\},\{T_{4},T_{13}\}, \{T_{8},T_{11}\},\{T_5,T_6,T_7,T_9,T_{10},T_{12},T_{14},T_{15}\}.$$    
For all branch lengths on this species tree, we also observe the inequalities 
\begin{equation}\label{E:inequalities3}
u_1>u_2,u_4,u_{8}>u_5
\end{equation}
and note that there are no other inequalities of the form $u_i > u_j$ that hold for all possible branch lengths.
In particular, the 8-element equivalence class of trees always has the strictly smallest probability associated with it.

\subsection{Species Tree Identifiability for 5 or more Taxa}
We will use several times the following observation, which is clear from the structure of the coalescent model. (In fact, this has already been used in Section \ref{sec:lininvcat} in the marginalization argument  explaining  a linear invariant for the caterpillar tree.)  While we state the lemma
for unrooted gene trees, there is of course a similar statement for the distribution of rooted gene trees.

\begin{lem} \label{lem:margin} If $S\subseteq X$ and $T'\in\mathcal T_S$, then
$$\mathbb P_{\sigma^+(S)}(T')= \sum_{T\in \mathcal T_{X} \atop T(S)=T'} \mathbb P_{\sigma^+}(T).$$
\end{lem}

As a consequence of the analysis for 4-taxon trees in Section \ref{sec:4taxa}, we obtain the following.

\begin{cor}\label{cor:unroot} For any $X$, $\mathbb P_{\sigma^+}$ determines $\sigma^-$.
\end{cor}
\begin{proof}
We assume $|X|\ge 4$, since otherwise there is nothing to prove.
For any quartet $Q\subseteq X$ of four distinct taxa, by Lemma \ref{lem:margin}, $\mathbb P_{\sigma^+}$ determines $\mathbb P_{\sigma^+(Q)}$. Thus $\sigma^-(Q)$ is determined by Proposition \ref{prop:4taxa}. Thus all unrooted quartet trees induced by $\psi^-$ are determined, along with their internal edge lengths. That all induced quartet topologies determine the topology $\psi^-$ is well known \citep{steel1992}. Because each internal edge of $\psi^-$ is the internal edge for some induced quartet tree, $\lambda^-$ is determined as well. \qed
\end{proof}

For the remaining arguments to determine $\sigma^+$, we may assume that $\sigma^-$ is already known.
We focus first on the $|X|=5$ case, and thus assume that $X=\{a,b,c,d,e\}$ and that $\psi^-$ has non-trivial splits $ab|cde$ and $abc|de$.

\begin{prop} \label{prop:cases}
For $|X|=5$ the rooted species tree topology $\psi^+$ is determined by  $\mathbb P_{\sigma^+}$.
\end{prop}
\begin{proof} 

From Section \ref{sec:lininv}, for generic values of $\lambda^+$, the caterpillar leads to seven distinct gene tree probabilities, with class sizes 1,1,1,2,2,2,6; the pseudocaterpillar gives five distinct probabilities, with class sizes 1,2,2,2,8; and the balanced tree gives five distinct probabilities, with class sizes 1,2,2,4,6. Thus the
(unlabeled) shape of $\psi^+$ can be distinguished for generic edge lengths. However, for certain values of these parameters the classes can degenerate, by merging.

To see that the tree shapes can be distinguished for all parameter values,  observe that the inequalities (\ref{E:inequalities1})--(\ref{E:inequalities3}) of Section \ref{sec:lininv} on gene tree probabilities  ensures the class associated to the smallest probability always has size 8 for the pseudocaterpillar, while for the other shapes the
size of this class is always 6. Moreover, for the caterpillar and balanced trees the size of the class associated to the second smallest probability must be exactly 2 and 4, respectively. Thus, these class sizes allow us to determine the unlabeled, rooted shape (balanced, caterpillar, or pseudocaterpillar) of the species tree.  In addition, from Corollary \ref{cor:unroot}, we also know the labeled, unrooted topology (i.e., the splits) of the species tree, $\psi^-$.  To determine the labeled, rooted topology, we consider cases depending on the unlabeled, rooted shape determined from the class sizes.

If the species tree is  balanced, from the splits we know that
$\psi^+=(((a,b),c),(d,e))$ or $\psi^+=((a,b),(c,(d,e)))$.
But the gene tree $T_7$, with splits $AD|BCE$ and $ABD|CE$, can be realized on the first of these species trees only if all coalescent events
occur above the root; on the second species tree, $T_7$ can be realized other ways as well. Thus $T_7$ would fall into the 6-element class of least probable gene trees for the first but not the second species tree. This then determines $\psi^+$.

For a caterpillar species tree, from the splits we know $\psi^+$ has as its unique 2-clade either $\{a,b\}$ or $\{d,e\}$. By considering the cherries on the two gene trees in the class of those with the second smallest probability, we see the 2-clade is determined as those taxa that appear in cherries with $c$. For simplicity, we henceforth suppose that the 2-clade is found to be $\{a,b\}$.  Thus, $\psi^+=((((a,b),c),d),e)$ or $((((a,b),c),e),d)$. 
Then from the inequality (\ref{E:inequalities2}) 
in Section  \ref{sec:lininvcat}, we find that $\mathbb P(T_3)>\mathbb P(T_2)$ if $\psi^+=((((a,b),c),d),e)$, while this inequality is reversed if $\psi^+=((((a,b),c),e),d)$.   

In the case of the pseudocaterpillar species tree, because the splits of $\psi^-$ are known, there is only one possibility for $\psi^+$.  Thus $\psi^+$ is determined.
\qed

\end{proof}

\begin{prop}\label{prop:5taxa}
For $|X|=5$, $\mathbb P_{\sigma^+}$ determines $\sigma^+=(\psi^+,\lambda^+)$.
\end{prop}

\begin{proof}
By Proposition \ref{prop:cases} , $\psi^+$ is determined.
From Corollary \ref{cor:unroot}, $\lambda^-$ is also determined.
Thus all elements of $\lambda^+$ except for the edges incident to the root are determined.
In the balanced case, the sum of these two unknown edge lengths is determined, but in the other cases we have yet to determine any information about the single such non-pendant edge length. We therefore consider each of these cases in order.  

If $\psi^+$ is balanced, we may assume $\sigma^+=(((a,b)\tc x,c)\tc y),(d,e)\tc z)$, with $y,z$ still to be determined. As the unrooted internal edge length $y+z$ is known, it is enough to determine $y$.
From the gene tree probabilities in Appendix \ref{app:gtdist1}, it follows that
\begin{align*}
XYZ&=6u_5+9u_7,\\
XY^3Z&=15u_7.
\end{align*}
Thus,
\begin{equation}\label{E:bl1}
y=-\log(Y)=\frac 12\log\left(\frac{2u_5+3u_7}{5u_7}\right ).
\end{equation}

If $\psi^+$ is a rooted caterpillar, we may assume $\sigma^+=((((a,b)\tc x,c)\tc y,d)\tc z, e)$. Only $z$ remains to be determined.
Using the explicit formulas for gene tree probabilities given in Appendix \ref{app:gtdist2}, one checks that
\begin{align*}
XY^3&= 3(-u_2+u_3+5u_7)\\
XY^3Z^6&=15(u_2-u_3+u_7)
\end{align*}
and thus
\begin{equation}\label{E:bl2}
z=-\log(Z)=\frac 16 \log\left (\frac {-u_2+u_3+5u_7}{5u_2-5u_3+5u_7} \right ).
\end{equation}

If  $\psi^+$ is the pseudocaterpillar, we may assume $\sigma^+=(((a,b)\tc x,(d,e)\tc y)\tc z,c)$, with $z$ still to be determined. The gene tree probabilities listed in  Appendix \ref{app:gtdist3} show that
\begin{align*}
XY&=12u_5+3u_8\\
XYZ^6&=30u_5-15u_8.
\end{align*}
Thus, 
\begin{equation}\label{E:bl3}
z=-\log(Z)=\frac 16\log\left( \frac{4u_5+u_8}{10u_5-5u_8}\right).
\end{equation}
Note that equations (\ref{E:bal1gtd})--(\ref{E:pcat1gtd})  of Appendix \ref{app:gtdist}
  can be used to show that the arguments of the logarithms in equations (\ref{E:bl1})--(\ref{E:bl3}) are always strictly greater than 1.
 \qed
\end{proof}

While this proof used particular formulas to identify the remaining edge lengths in $\lambda^+$, note that many variants could have been used in their place. This simply reflects the many algebraic relationships (both linear and non-linear invariants) between the various gene tree probabilities.

\smallskip

With the $|X|=5$ case completed, we obtain the general result.

\begin{thm} \label{thm:main}The unrooted topological gene tree distribution $\mathbb P_{\sigma^+}$ arising from the multspecies coalescent model for samples of one lineage per taxon determines the metric species tree $\sigma^+$ provided $|X|\ge 5$. If $|X|=4$, $\mathbb P_{\sigma^+}$ determines only the unrooted metric species tree $\sigma^-$.
\end{thm}
\begin{proof} By Corollary \ref{cor:unroot}, $\sigma^-=(\psi^-,\lambda^-)$ is determined. 
 
If $|X|\ge 5$, consider a specific edge $e$ of $\psi^-$, and all 5-taxon subsets $S\subseteq X$ such that the induced unrooted tree $\psi^-(S)$ has $e$ as an edge. If the root, $\rho$, of $\psi^+$  lies on $e$, then the root of $\psi^+(S)$ is also $\rho$ and thus the root of $\psi^+(S)$ lies on $e$ for all such $S$. If $\rho$ does not lie on $e$, then there exists an $S$ with the root of $\psi^+(S)$ not on $e$. To see this, for any set $Q\subset X$ of four taxa which distinguishes $e$ \citep[Proposition 6]{steel1992}, choose $x \in X\setminus Q$ 
so that the MRCA of $S= Q \cup \{x\}$ is $\rho$. Then $\psi^+(S)$ has root $\rho$, which is not on $e$.

Thus using
Lemma \ref{lem:margin} and Proposition \ref{prop:5taxa} to determine the root location of  such $\psi^+(S)$ for each $e$, we can determine $\psi^+$. Then the length of any internal edges incident to the root of $\psi^+$ can be recovered by choosing a 5-taxon subset $S$ such that $\psi^+(S)$ has these edges, and applying Lemma \ref{lem:margin} and Proposition \ref{prop:5taxa} again.
Thus $\sigma^+$ is determined.

Proposition \ref{prop:4taxa} gives the case $|X|=4$. \qed
\end{proof}

Theorem \ref{thm:main}  gives an alternate approach to establishing  Corollary \ref{cor:rgt}, in cases with $|X|\ge 5$, since the distribution of rooted gene trees determines that of unrooted gene trees.

Theorem \ref{thm:main} can also be used to show that if multiple lineages are sampled from some or all of the taxa, then the unrooted gene tree distribution contains additional information on the species tree, as follows.

\begin{cor}\label{cor:intra} Consider a species tree on taxon set $X$,  and, for some $\ell_i>0$, the distribution of unrooted topological gene trees under a multispecies coalescent model of samples of $\ell_i$ individuals from taxon $i$. Suppose that either $|X|\ge 4$ and that there is at least one $i$ such that $\ell_i\ge 2$, or that  $|X|=3$ and that there are at least two values of $i$ such that $\ell_i\ge 2$.
Then the gene tree distribution determines the species tree's rooted topology,  internal edge lengths, and for any taxon with $l_i>1$ the length of the pendant edge leading to taxon $i$.
\end{cor}

\begin{proof}
We may assume all $\ell_i$ are either 1 or 2, by marginalizing over any additional individuals sampled, if necessary.

Construct an extended species tree by attaching to any leaf $i$ for which $l_i=2$ a pair of edges descending to new leaves labelled $i_1$ and $i_2$, so the extended species tree has $\ell = \sum_{i=1}^n \ell_i$ leaves. The pendant edge leading to taxon $i$ in the original species tree becomes an internal edge on the extended tree, and retains its previous length. The lengths of the new pendant edges in the extended tree can be chosen arbitrarily, or left unspecified. Then a coalescent process on the extended $\ell$-taxon tree with one sample per leaf leads to exactly the same distribution of topological gene trees as the  the multiple-sample process on the original species tree.

Applying Theorem \ref{thm:main} to the extended tree, we obtain the result. \qed
\end{proof}

\section{Nonbinary species trees}\label{sec:nonbinary}

The results for binary species trees generalize to nonbinary species trees as well.  
When species trees are allowed to be nonbinary, there are two unlabeled 3-taxon  tree shapes, five  unlabeled 4-taxon tree shapes, and 12  unlabeled 5-taxon tree shapes \citep{cayley1857}. 
Probabilities of binary, unrooted gene tree topologies given a nonbinary species 
tree can be obtained by considering the limiting probability as one or more branch lengths go to zero in the formulas derived for binary species trees.  We note that under the standard Kingman coalescent, gene trees, which depend on exponential waiting times, are still binary with probability 1 even when the species tree has polytomies.

For the $3$-taxon species tree, $((a,b)\tc t,c)$, letting $t \rightarrow 0$, the rooted gene 
tree probabilities are each 1/3 in the limit.  Thus the unresolved 3-taxon rooted species tree 
can be identified from the gene tree distribution from the presence of three equal probabilities; 
whereas for a resolved species tree, exactly one gene tree has probability greater than 1/3. 
Similarly, polytomies in any larger species tree can be identified by considering rooted 
triplets.  A species tree node has three or more descendants if the three rooted gene trees 
obtained from sampling one gene from three distinct descendants of the node have equal probabilities.  

For $4$-taxon species trees, the completely unresolved topology $(a,b,c,d)$ can not be distinguished from the partially unresolved $((a,b,c)\tc y,d)$ from unrooted gene tree probabilities as both result in equal probabilities of the three binary, unrooted gene trees on these taxa. Similarly, the resolved species trees $(((a,b)\tc x,c)\tc y,d)$ and the partially unresolved  $((a,b)\tc x, c, d)$ yield the same unrooted gene tree probabilities, with $\mathbb P_{\sigma^+}(\mathcal{T}_{AB|CD}) = 1 - \frac{2}{3}e^{-x}$.   
These observations lead to the conclusion, as in the binary case, that $
4$-taxon unrooted gene tree probabilities identify the unrooted (possibly unresolved) species tree, but do not identify the root.  Thus Proposition \ref{prop:4taxa} is still valid when  $\sigma^+$ is nonbinary.
  
Identifiability of possibly-nonbinary rooted species trees for 5 or more taxa from probabilities of unrooted gene tree topologies  can be established using arguments similar to those of the binary case.  While we defer the detailed proofs  to Appendix  \ref{sec:app-nonbinary}, we state these results as follows:

\begin{prop}\label{prop:nonbinary}
Proposition \ref{prop:4taxa}, Corollary \ref{cor:unroot}, 
Propositions  \ref{prop:cases} and  \ref{prop:5taxa}, 
Theorem \ref{thm:main}, and Corollary \ref{cor:intra} remain valid if $\sigma^+$ is nonbinary.
\end{prop}

We note that a species tree with a polytomy is equivalent to a model of a resolved species tree with one or more branch lengths set equal to zero, and therefore that a resolved species tree and a polytomous species tree can be regarded as nested models.  Although it might be difficult to distinguish  polytomous versus resolved species trees from finite amounts of data, the nested relationship of these models suggests that likelihood ratios could be used to determine whether an estimated species tree branch length is significantly greater than 0.  A previous study \citep{poe2004} argued for the hypothesis of a species-level polytomy in early bird evolution by using likelihood ratios to test whether gene trees had branch lengths significantly greater than 0 at multiple loci.  Since gene trees are theoretically expected to be resolved under the coalescent model, an alternative procedure would be to use probabilities of gene trees under the polytomous and resolved species trees and perform a likelihood ratio test for whether an estimated species tree branch length is significantly greater than 0.

\section{Discussion}\label{sec:Discuss}

Under standard models of sequence evolution, the distribution of site patterns of DNA does not depend on the position of the root of the gene tree on which the sequence evolve (this is sometimes called the ``pulley principle'' \citep{felsenstein1981}).  Inference of the root of a gene tree requires
additional assumptions, such as that of a molecular clock (mutation occurs at a constant rate throughout the tree), or inclusion of an outgroup taxon in the analysis, so that the root may be assumed to lie where the outgroup joins all other taxa in the study.   We have shown, however, that under the coalescent model with five  or more species, the distribution of unrooted topological gene trees preserves information about both the rooted species tree and its internal branch lengths.   Thus in multilocus studies in which many gene trees are inferred, it is theoretically possible to infer the rooted metric species tree even in the absence of a molecular clock, known outgroups, or any metric information on the gene trees. While for some data sets it can be difficult to obtain either reliable roots or evolutionary times for branches of gene trees, these issues are not fundamental barriers to species tree inference.

Although we have shown the theoretical possibility of identifying rooted species trees from unrooted gene trees by using linear invariants, we emphasize that we do not propose using these invariants as a basis for inference.
Invariants of gene tree distributions are functions of their exact probabilities under the model --- from finite data sets, gene trees are inferred with some error, and empirical estimates of gene tree probabilities from a finite number of gene trees might not satisfy invariants or inequalities that apply to the exact distribution.   
Moreover, many non-linear invariants which are not discussed in this paper (and not yet fully understood) further constrain the form of the gene tree distribution. 

In practice, very large numbers of loci might be needed to obtain approximate estimates of gene tree probabilities, and there must be considerable gene tree discordance in order to estimate probabilities of less probable unrooted gene trees.  For example, in an often-analyzed 106-gene yeast dataset \citep{rokas2003}, analyzing only the five species about which there is the most conflict, ({\it S.~cerevisiae}, {\it S.~paradoxus}, {\it S.~mikatae}, {\it S.~kudriavzevil}, and {\it S.~bayanus}), yields the same unrooted gene tree for all 106 loci when inferred using maximum likelihood under the $\text{GTR}+\Gamma+\text{I}$ model without a molecular clock.  If all observed gene trees have the same unrooted topology, then there is not enough information to infer the rooted species tree.
 Other data have shown more conflict in unrooted gene trees, such as a 162-gene dataset for rice \citep{cranston2009}, in which 99 of 105 rooted 5-taxon gene trees were represented in the Bayesian 95\% highest posterior density (HDP) set of trees.\\

If species tree branches are too long, it will not be possible to recover the rooted species tree from finite data.  For example, if the species tree is $(((a,b)\tc x,c)\tc y,d)\tc z,e)$, where $y$ is sufficiently large, every observed gene tree (for a finite number of loci) might have the $ABC|DE$ split.  Being able to determine that $e$ is the outgroup would require observing conflicting splits, such as that $ABD|CE$ is more probable than $ABE|CD$.  However, if $y$ is large, these conflicting splits are likely to never be observed, making it difficult to distinguish between rooted topologies $(((a,b),c),d),e)$, $(((a,b),c),e),d)$, and $(((a,b),c),(d,e))$.
 
 On the other hand, if branches are too short, it might be difficult to distinguish between certain rooted species trees, such as between $(((a,b),c),(d,e))$ and $((a,b),(c,(d,e)))$ when the node immediately ancestral to $c$ is very close to the root.  Further study would be needed for a precise understanding of how extreme branch lengths affect the number of gene trees needed for reliable inference of the species tree. We note, however, that even when the rooted species tree cannot be fully inferred with great certainty, some rooted aspects of the tree might be recoverable. For example in the case of the rooted trees above, one might infer that $(a,b)$ and $(d,e)$ are rooted cherries on the tree, even if the placement of taxon $c$ with respect to the root remains unknown.
 
We again emphasize that invariants are not the most promising approach for inferring species trees from finite data, and that a maximum likelihood (ML)  or Bayesian approach might be more appropriate.
Given a set of sufficiently conflicting unrooted topological gene trees inferred by standard methods and then assumed to be correct, the rooted species tree could be inferred using ML, where  the likelihood of the species tree is 
\begin{equation}\label{E:ML}
L(\sigma^+) \propto \prod_{i=1}^{(2n-5)!!} u_i^{n_i}
\end{equation}
where there are $n$ taxa and  the $i$th unrooted gene tree topology is observed $n_i$ times with $\sum_i n_i = N$ the total number of loci. The probability $u_i$ of the $i$th gene tree depends on the species tree topology and branch lengths as outlined in Section \ref{sec:coalmodel}.  However, this 2-stage approach of gene tree inference followed by species tree inference does not take into account uncertainty in the gene trees, or cases in which inferred gene trees are not fully resolved.
If there is not enough information in the sequences to estimate resolved gene trees, an approximation to  equation (\ref{E:ML}) would be to either randomly resolve the tree if there are very many loci  (as is often done in software implementing quartet puzzling \citep{StrimmerAndvonHaeseler96} or neighbor joining \citep{saitou.nei:nj}); or, if an unresolved gene tree has $k$ resolutions, let the  locus contribute a count of $1/k$ to each resolution.

To better utilize the information in the unrooted gene trees, an attractive, but computationally more intensive, approach would use a Bayesian framework in which the posterior distribution of the rooted species tree is determined from posterior distributions of gene trees, thus
taking into account uncertainty in the estimated gene trees. 
Cases in which ML would return an unresolved gene tree would likely correspond to a posterior distribution of gene trees with substantial support on more than one topology. Thus, instead of each locus contributing a count of one gene tree topology, it contributes fractional proportions to several topologies.
In cases in which the gene tree distributions carry little information about the root of the species tree,  the posterior distribution of the species tree would indicate this uncertainty by spreading the posterior mass over several species 
trees.  The results of the present paper  suggest that it is possible to extend current model-based methods of inferring rooted species trees (e.g., BEST \citep{liu2007} and STEM \citep{kubatko2009}) to cases where only unrooted gene trees can be estimated.

Finally we note that invariants have a potential use in testing the fit of the multispecies coalescent model to a dataset.  As noted in \citep{Slatkin2008}, processes such as population subdivision can lead to asymmetry in the probabilities of the two nonmatching rooted gene trees in the case of three taxa, thus violating the invariant in equation (\ref{E:invariant3taxa}).  As shown in this paper, similar invariants can be obtained for larger number of species even when only unrooted gene trees are available, thus allowing the testing of the fit of the multispecies coalescent model in situations more general than the rooted 3-taxon setting.

\begin{acknowledgements}
The authors thank the Statistical and Applied Mathematical Sciences Institute, where this work was begun during its 2008-09 program on Algebraic Methods in Systems Biology and Statistics. We also thank two anonymous reviewers, one of whom suggested the extension to nonbinary trees.   ESA and JAR were supported by funds from the National Science Foundation,  grant DMS 0714830, and JAR by an Erskine Fellowship from the University of Canterbury. JHD was funded by the New Zealand Marsden Fund. All authors contributed equally to this work.
\end{acknowledgements}

\begin{footnotesize}
\bibliographystyle{spbasic}
\bibliography{bibfile}
\end{footnotesize}

\newpage
\appendix
\section{Tables for 5-taxon trees }\label{app:tables}
\begin{table}[!h]
\caption{The 105 rooted gene trees on 5 species.}
\label{T:5taxarooted}
\begin{center}
\begin{tabular}{l c l c l c}
\hline
$R_1$	&	$((((A,B),C),D),E)$   		&	$R_{36}$	&	$((((B,D),E),C),A)$     	&	$R_{71}$	&	$(((A,D),(C,E)),B)$\\ 
$R_2$	&	$((((A,B),C),E),D)$		&	$R_{37}$	&	$((((B,E),A),C),D)$		&	$R_{72}$	&	$(((A,E),(C,D)),B)$\\ 
$R_3$	&	$((((A,B),D),C),E)$		&	$R_{38}$	&	$((((B,E),A),D),C)$		&	$R_{73}$	&	$(((B,C),(D,E)),A)$\\ 
$R_4$	&	$((((A,B),D),E),C)$		&	$R_{39}$	&	$((((B,E),C),A),D)$		&	$R_{74}$	&	$(((B,D),(C,E)),A)$\\ 
$R_5$	&	$((((A,B),E),C),D)$		&	$R_{40}$	&	$((((B,E),C),D),A)$		&	$R_{75}$	&	$(((B,E),(C,D)),A)$\\ 
$R_6$      &       $((((A,B),E),D),C)$             &       $R_{41}$ &       $((((B,E),D),A),C)$             &       $R_{76}$	&	$(((A,B),C),(D,E))$\\
$R_7$	&	$((((A,C),B),D),E)$		&	$R_{42}$	&	$((((B,E),D),C),A)$		&	$R_{77}$	&	$(((A,C),B),(D,E))$\\ 
$R_8$	&	$((((A,C),B),E),D)$		&	$R_{43}$	&	$((((C,D),A),B),E)$		&	$R_{78}$	&	$(((B,C),A),(D,E))$\\ 
$R_9$	&	$((((A,C),D),B),E)$		&	$R_{44}$	&	$((((C,D),A),E),B)$		&	$R_{79}$	&	$(((A,B),D),(C,E))$\\ 
$R_{10}$	&	$((((A,C),D),E),B)$		&	$R_{45}$	&	$((((C,D),B),A),E)$		&	$R_{80}$	&	$(((A,D),B),(C,E))$\\ 
$R_{11}$	&	$((((A,C),E),B),D)$		&	$R_{46}$	&	$((((C,D),B),E),A)$		&	$R_{81}$	&	$(((B,D),A),(C,E))$\\ 
$R_{12}$	&	$((((A,C),E),D),B)$		&	$R_{47}$	&	$((((C,D),E),A),B)$		&	$R_{82}$	&	$(((A,C),D),(B,E))$\\ 
$R_{13}$	&	$((((A,D),B),C),E)$		&	$R_{48}$	&	$((((C,D),E),B),A)$		&	$R_{83}$	&	$(((A,D),C),(B,E))$\\ 
$R_{14}$	&	$((((A,D),B),E),C)$		&	$R_{49}$	&	$((((C,E),A),B),D)$		&	$R_{84}$	&	$(((C,D),A),(B,E))$\\ 
$R_{15}$	&	$((((A,D),C),B),E)$		&	$R_{50}$	&	$((((C,E),A),D),B)$		&	$R_{85}$	&	$(((B,C),D),(A,E))$\\ 
$R_{16}$	&	$((((A,D),C),E),B)$		&	$R_{51}$	&	$((((C,E),B),A),D)$		&	$R_{86}$	&	$(((B,D),C),(A,E))$\\ 
$R_{17}$	&	$((((A,D),E),B),C)$		&	$R_{52}$	&	$((((C,E),B),D),A)$		&	$R_{87}$	&	$(((C,D),B),(A,E))$\\ 
$R_{18}$	&	$((((A,D),E),C),B)$		&	$R_{53}$ &       $((((C,E),D),A),B)$             &       $R_{88}$ &       $(((A,B),E),(C,D))$\\
$R_{19}$  &      $((((A,E),B),C),D)$             &       $R_{54}$	&	$((((C,E),D),B),A)$		&	$R_{89}$	&	$(((A,E),B),(C,D))$\\ 
$R_{20}$	&	$((((A,E),B),D),C)$		&	$R_{55}$	&	$((((D,E),A),B),C)$		&	$R_{90}$	&	$(((B,E),A),(C,D))$\\ 
$R_{21}$	&	$((((A,E),C),B),D)$     	&	$R_{56}$ &       $((((D,E),A),C),B)$             &       $R_{91}$ &       $(((A,C),E),(B,D))$\\		 
$R_{22}$ &       $((((A,E),C),D),B)$             &       $R_{57}$	&	$((((D,E),B),A),C)$             &	$R_{92}$	&	$(((A,E),C),(B,D))$\\
$R_{23}$	&	$((((A,E),D),B),C)$     	&	$R_{58}$	&	$((((D,E),B),C),A)$		&       $R_{93}$ &       $(((C,E),A),(B,D))$\\ 
$R_{24}$ &       $((((A,E),D),C),B)$             &       $R_{59}$ &       $((((D,E),C),A),B)$             &	$R_{94}$	&	$(((B,C),E),(A,D))$\\
$R_{25}$	&	$((((B,C),A),D),E)$		&	$R_{60}$	&	$((((D,E),C),B),A)$		&	$R_{95}$	&	$(((B,E),C),(A,D))$\\ 
$R_{26}$	&	$((((B,C),A),E),D)$		&	$R_{61}$	&	$(((A,B),(C,D)),E)$		&	$R_{96}$	&	$(((C,E),B),(A,D))$\\ 
$R_{27}$	&	$((((B,C),D),A),E)$		&	$R_{62}$	&	$(((A,C),(B,D)),E)$		&	$R_{97}$	&	$(((A,D),E),(B,C))$\\ 
$R_{28}$	&	$((((B,C),D),E),A)$		&	$R_{63}$	&	$(((A,D),(B,C)),E)$		&       $R_{98}$ &       $(((A,E),D),(B,C))$\\
$R_{29}$ &       $((((B,C),E),A),D)$             &       $R_{64}$ &       $(((A,B),(C,E)),D)$             &	$R_{99}$	&	$(((D,E),A),(B,C))$\\
$R_{30}$	&	$((((B,C),E),D),A)$		&	$R_{65}$	&	$(((A,C),(B,E)),D)$		&	$R_{100}$ &	$(((B,D),E),(A,C))$\\ 
$R_{31}$	&	$((((B,D),A),C),E)$		&	$R_{66}$	&	$(((A,E),(B,C)),D)$		&	$R_{101}$	&	$(((B,E),D),(A,C))$\\ 
$R_{32}$	&	$((((B,D),A),E),C)$		&	$R_{67}$	&	$(((A,B),(D,E)),C)$		&	$R_{102}$	&	$(((D,E),B),(A,C))$\\ 
$R_{33}$	&	$((((B,D),C),A),E)$		&	$R_{68}$	&	$(((A,D),(B,E)),C)$		&	$R_{103}$	&	$(((C,D),E),(A,B))$\\ 
$R_{34}$	&	$((((B,D),C),E),A)$		&	$R_{69}$	&	$(((A,E),(B,D)),C)$		&	$R_{104}$	&	$(((C,E),D),(A,B))$\\ 
$R_{35}$	&	$((((B,D),E),A),C)$		&	$R_{70}$	&	$(((A,C),(D,E)),B)$		&	$R_{105}$	&	$(((D,E),C),(A,B))$\\ 
\hline
\end{tabular}
\end{center}
\end{table}


\begin{table}[b]
\caption{The 15 unrooted 5-taxon topological gene trees, as indicated by their non-trivial splits, and their probabilities $u_i=\mathbb P(T_i)$, where $r_i$ is the probability of the rooted gene tree $R_i$ given the species tree $\sigma^+$.}
\label{T:5taxaprobs}
\begin{center}
\begin{tabular}{l l l } 
Tree & \multicolumn{1}{c}{Splits} &\multicolumn{1}{c}{Probability}\\
\hline
$T_1$ & $AB|CDE$, $ABC|DE$ & $u_1=  r_1 + r_2+r_{59}+r_{60}+r_{67}+r_{76}+r_{105}$ \\                                                                              
$T_2$ & $AB|CDE$, $ABD|CE$ & $u_2=  r_3 + r_4+r_{53}+r_{54}+r_{64}+r_{79}+r_{104}$ \\                                                                               
$T_3$ & $AB|CDE$, $ABE|CD$ & $u_3=  r_5 + r_6+r_{47}+r_{48}+r_{61}+r_{88}+r_{103}$ \\                                                                            
$T_4$ & $AC|BDE$, $ABC|DE$ & $u_4=  r_7 + r_8+r_{57}+r_{58}+r_{70}+r_{77}+r_{102}$  \\                                                                              
$T_5$ & $AC|BDE$, $ACD|BE$ & $u_5=  r_9 +r_{10}+r_{41}+r_{42}+r_{65}+r_{82}+r_{101}$  \\                                                                               
$T_6$ & $AC|BDE$, $ACE|BD$ & $u_6=  r_{11}+r_{12}+r_{35}+r_{36}+r_{62}+r_{91}+r_{100}$ \\                                                                           
$T_7$ & $AD|BCE$, $ABD|CE$ & $u_7=  r_{13}+r_{14}+r_{51}+r_{52}+r_{71}+r_{80}+r_{96}$\\                                                                              
$T_8$ & $AD|BCE$, $ACD|BE$ & $u_8=  r_{15}+r_{16}+r_{39}+r_{40}+r_{68}+r_{83}+r_{95}$\\                                                                             
$T_9$ & $AD|BCE$, $ADE|BC$ & $u_9=  r_{17}+r_{18}+r_{29}+r_{30}+r_{63}+r_{94}+r_{97}$  \\                                                                          
$T_{10}$ & $AE|BCD$, $ABE|CD$ & $u_{10}=  r_{19}+r_{20}+r_{45}+r_{46}+r_{72}+r_{87}+r_{89}$  \\                                                                               
$T_{11}$ & $AE|BCD$, $ACE|BD$ & $u_{11}=  r_{21}+r_{22}+r_{33}+r_{34}+r_{69}+r_{86}+r_{92}$  \\                                                                             
$T_{12}$ & $AE|BCD$, $ADE|BC$ & $u_{12}=  r_{23}+r_{24}+r_{27}+r_{28}+r_{66}+r_{85}+r_{98}$  \\                                                                              
$T_{13}$ & $BC|ADE$, $ABC|DE$ & $u_{13}= r_{25}+r_{26}+r_{55}+r_{56}+r_{73}+r_{78}+r_{99}$  \\                                                                              
$T_{14}$ & $BD|ACE$, $ABD|CE$ & $u_{14}=  r_{31}+r_{32}+r_{49}+r_{50}+r_{74}+r_{81}+r_{93}$  \\                                                                             
$T_{15}$ & $BE|ACD$, $ABE|CD$ & $u_{15}=  r_{37}+r_{38}+r_{43}+r_{44}+r_{75}+r_{84}+r_{90}$ \\
\hline  
\end{tabular}
\end{center}
\end{table}

\clearpage

\section{5-taxon unrooted gene tree distributions}\label{app:gtdist}

\subsection{Balanced species tree}\label{app:gtdist1}
For the 5-taxon balanced species tree of Fig.~\ref{F:modeltrees}c,
$$\sigma^+=(((a,b)\tc x,c)\tc y,(d,e)\tc z),$$  let $X=\exp(-x)$, $Y=\exp(-y)$, and $Z=\exp(-z)$. Then the distribution of
unrooted gene trees $T_i$ is given by $u_i=\mathbb P_{\sigma^+}(T_i)$ with
\begin{align}\label{E:bal1gtd}
u_1 &= 1-\frac{2}{3}X-\frac{2}{3}YZ+\frac{1}{3}XYZ+\frac{1}{15}XY^3Z,\notag\\
u_2 &= u_3 = \frac{1}{3}YZ-\frac{1}{6}XYZ-\frac{1}{10}XY^3Z,\notag\\
u_4 &= u_{13} = \frac{1}{3}X -\frac{1}{3}XYZ+\frac{1}{15}XY^3Z,\notag\\
u_5 &= u_6 = u_9 = u_{12} = \frac{1}{6}XYZ -\frac{1}{10}XY^3Z,\notag\\
u_7 &= u_8 = u_{10} = u_{11} = u_{14} = u_{15} = \frac{1}{15}XY^3Z.
\end{align}



\subsection{Rooted caterpillar species tree}\label{app:gtdist2}
For the 5-taxon rooted caterpillar species tree of Fig.~\ref{F:modeltrees}d,
$$\sigma^+=((((a,b)\tc x,c)\tc y,d)\tc z, e),$$ let $X=\exp(-x)$, $Y=\exp(-y)$, and $Z=\exp(-z)$. Then the distribution of
unrooted gene trees $T_i$ under the coalescent is given by $u_i=\mathbb P_{\sigma^+}(T_i)$ with
\begin{align}\label{E:cat1gtd}
u_1 &= 1-\frac{2}{3}X-\frac{2}{3}Y+\frac{1}{3}XY+\frac{1}{18}XY^3+\frac{1}{90}XY^3Z^6,\notag\\
u_2 &= \frac{1}{3}Y-\frac{1}{6}XY -\frac{1}{9}XY^3 +\frac{1}{90}XY^3Z^6,\notag\\
u_3 &= \frac{1}{3}Y-\frac{1}{6}XY-\frac{1}{18}XY^3-\frac{2}{45}XY^3Z^6, \notag\\ 
u_4 &= u_{13} = \frac{1}{3}X -\frac{1}{3}XY + \frac{1}{18}XY^3+\frac{1}{90}XY^3Z^6,\notag \\
u_5 &= u_{12} = \frac{1}{6}XY - \frac{1}{9}XY^3+\frac{1}{90}XY^3Z^6,\notag\\
u_6 &= u_9 = \frac{1}{6}XY  - \frac{1}{18}XY^3 -\frac{2}{45}XY^3Z^6,\notag\\
u_7 &= u_8 = u_{10} = u_{11} = u_{14} = u_{15} = \frac{1}{18}XY^3+ \frac{1}{90}XY^3Z^6.
\end{align}




\subsection{Pseudocaterpillar species tree}\label{app:gtdist3}
For the 5-taxon pseudocaterpillar species tree of Fig.~\ref{F:modeltrees}e, $$\sigma^+=(((a,b)\tc x,(d,e)\tc y)\tc z,c),$$ let $X=\exp(-x)$, $Y=\exp(-y)$, and $Z=\exp(-z)$. Then the distribution of
unrooted gene trees $T_i$ is given by $u_i=\mathbb P_{\sigma^+}(T_i)$ with

\begin{align}\label{E:pcat1gtd}
u_1 &= 1-\frac{2}{3}X - \frac{2}{3}Y +\frac{4}{9}XY - \frac{2}{45}XYZ^6,\notag\\
u_2 &= u_3 = \frac{1}{3}Y-\frac{5}{18}XY+\frac{1}{90}XYZ^6,\notag\\
u_4 &= u_{13} = \frac{1}{3}X-\frac{5}{18}XY+\frac{1}{90}XYZ^6,\notag\\  
u_5 &= u_6 = u_7 = u_9 = u_{10} = u_{12} = u_{14} = u_{15} = \frac{1}{18}XY+\frac{1}{90}XYZ^6,\notag\\
u_8 &= u_{11} = \frac{1}{9}XY - \frac{2}{45}XYZ^6.
\end{align}

\section{Nonbinary species trees}\label{sec:app-nonbinary}

\begin{proof}[of Proposition \ref{prop:nonbinary}] 
The extension of Proposition \ref{prop:4taxa} to nonbinary $\sigma^+$ was discussed in Section \ref{sec:nonbinary}.

From this, for $|X|\ge5$ we know that for $Q \subset X$ with $|Q| = 4$, the possibly unresolved unrooted quartet tree on $Q$ can be determined from gene tree probabilities. Thus  the unrooted, labeled species tree $\sigma^-$ can be determined by the identifiability of (possibly nonbinary) phylogenetic trees from their quartets \citep{bandelt1986}\citep[Theorem 6.3.5]{semple2003}, and thus Corollary \ref{cor:unroot} has been extended.

 Next, in addition to the three fully resolved rooted tree shapes on 5 taxa, we must consider the nine rooted shapes with polytomies. In Table \ref{T:poly1}, we designate these
 as $P_1, \dots, P_9$, specify an arbitrary labeling of the leaves of each, and list inequalities analogous to inequalities (\ref{E:inequalities1})--(\ref{E:inequalities3}) for unrooted gene tree probabilities.  The equivalence classes of labeled, binary, unrooted 5-taxon gene trees associated with each polytomous species tree are given in Table \ref{T:poly2}, along with the gene tree probabilities as functions of  transformed branch lengths $X$, $Y$, and $Z$.  Gene tree probabilities are obtained from the equations for resolved trees in Appendix \ref{app:gtdist} by setting one or more branch lengths to 0.

For the 5-taxon species tree shapes, in all cases of either resolved and polytomous trees, the least probable class, $\mathcal C$, of gene trees always has probability strictly smaller than all others. There are five possible cases for the cardinality of $\mathcal C$:

\begin{enumerate}
\item $|\mathcal C|=15$:  polytomy $P_1$
\item $|\mathcal C|=12$: polytomy $P_2$ or polytomy $P_3$
\item $|\mathcal C|=10$:  polytomy $P_5$ or polytomy $P_7$
\item$|\mathcal C|=8$: resolved pseudocaterpillar
\item $|\mathcal C|=6$: resolved caterpillar, resolved balanced,
 polytomy $P_4$, polytomy $P_6$,  polytomy $P_8$, or polytomy $P_9$
\end{enumerate}

If $|\mathcal C|=12$, we can distinguish between $P_2$ and $P_3$ since all gene trees in the 3-element class for $P_3$ 
have the same taxon not occurring in a cherry, while for $P_2$ the gene trees in the 3-element class have different taxa in this role.

For $|\mathcal C|=10$, we can distinguish between polytomies $P_5$ and $P_7$ by considering the two 2-element classes for both. For $P_5$,
both of the classes $\{T_2,T_3\}$ and  $\{T_4,T_{13}\}$ contain trees with one cherry in common.
For $P_7$,
the gene trees in the classes $\{T_2,T_3\}$ have a cherry in common, but those in  $\{T_8,T_{11}\}$ do not. 
Note that it is possible for these classes to degenerate, to form a 4-element class, but by counting the number of trees with a cherry in common in the larger degenerate class we can still determine whether the species tree
shape is $P_5$ or $P_7$.

If $|\mathcal C|=6$ the cardinality of the class with the second smallest probability determines the rooted tree shape in some cases. The class with the second smallest probability has cardinality 2 only for the resolved caterpillar, cardinality 3 only for $P_9$, cardinality 6 only for polytomies $P_4$ and $P_8$, and cardinality 4 for the resolved balanced tree and $P_6$.

At this point we have determined the rooted unlabeled topology of the species tree from the 5-taxon gene tree classes, except for the $P_4$ versus $P_8$ case and the balanced versus $P_6$ case. (We will return to these cases later.)\\

For the fully resolved trees, Proposition \ref{prop:5taxa} explains how we determine the labeling, so similar arguments are needed for each polytomous tree. 
If the species tree is $P_1$, there is nothing to do.
For polytomies $P_2$, $P_5$, and $P_4$/$P_8$ the labeling on the unrooted tree determines that on the rooted one.

If the species tree is $P_3$, the taxon that appears in no cherry in the gene trees in the 3-element class is the one that is an outgroup to all the others in the species tree.

For polytomy $P_7$, the resolved cherry in the species tree is 
determined by the unrooted labeled tree, and the 
outgroup is determined by not appearing in a cherry in the most probable gene tree. 

For polytomy $P_9$, the non-outgroup taxon which is not descended from the polytomy in the species tree is distinguished by not appearing in any cherry in the class with the second smallest probability. Calling this identified taxon $d$, the outgroup taxon is determined as the one appearing in a cherry with $d$ 
in three of the six most probable trees (i.e., in three trees in the union of the two most probable classes, which may degenerate to a single class).

 Finally, the labeling on the balanced/$P_6$ tree is determined as  it was for the balanced tree in the proof of Proposition \ref{prop:cases}.

At this point we have determined the labeled rooted species tree topology $\psi^+$ (except for two $P_4$/$P_8$ and balanced/$P_6$ ambiguities).\\

It remains to determine branch lengths on $\psi^+$.   If the unlabeled species tree is any tree other than the resolved balanced tree, $P_4$, $P_6$, or $P_8$, then the branch lengths can be solved from the system of equations listed for the given species tree from Appendix \ref{app:gtdist} or the formulas in Table \ref{T:poly2}.   

If the species tree is known to be either $P_4$ or $P_8$, we note that $P_8$ degenerates to $P_4$ as $z \rightarrow 0$ (or $Z \rightarrow 1$).  Solving the system of equations for the $u_i$s in terms of the branch lengths for $P_8$, if $Z < 1$, then the species tree is $P_8$.  If $Z=1$, then the species tree is $P_4$.
 Similarly, $P_6$ is the limiting case of  the balanced 5-taxon species tree as $Z \rightarrow1$.  Solving the system of equations for the balanced tree, the species tree is the balanced tree if $Z < 1$ and is $P_6$ if $Z=1$.

Thus,  for a 5-taxon species  tree, even with polytomies, $\sigma^+$ is identifiable.  This extends Propositions \ref{prop:cases} and \ref{prop:5taxa} to potentially nonbinary species trees.  Theorem \ref{thm:main} also extends, noting that if the root of the species tree has degree greater than 2, then its location will be identified by some 5-taxon subtree with the same property. 
The proof of Corollary \ref{cor:intra} did not use the assumption that $\sigma^+$ is binary, so it applies to nonbinary species trees as well. 
\qed 
\end{proof}

\begin{table}
 \begin{center}
 \caption{Representatives for the 9 nonbinary, rooted 5-taxon species tree shapes, with inequalities for gene tree probabilities.
 }\label{T:poly1}
 \begin{tabular}{l l l c}
Species tree & Newick representative & inequalities & species tree shape\\
\hline $P_1$ &   $(a,b,c,d,e)$ & &  \includegraphics[width=.88cm]{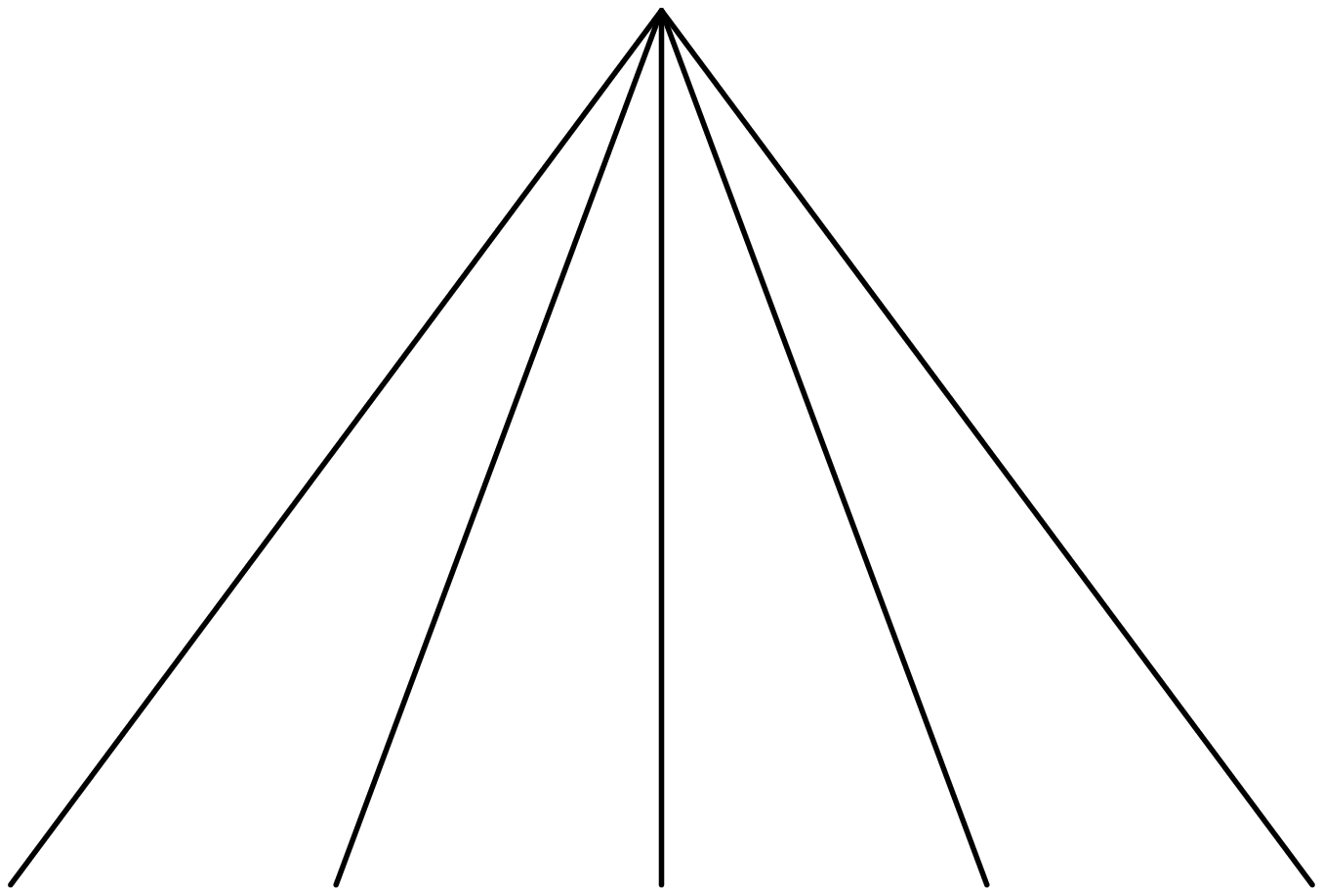}   \\
$P_2$  & $(a,b,c,(d,e)\tc z)$ & $u_1 > u_2$ & \includegraphics[width=.88cm]{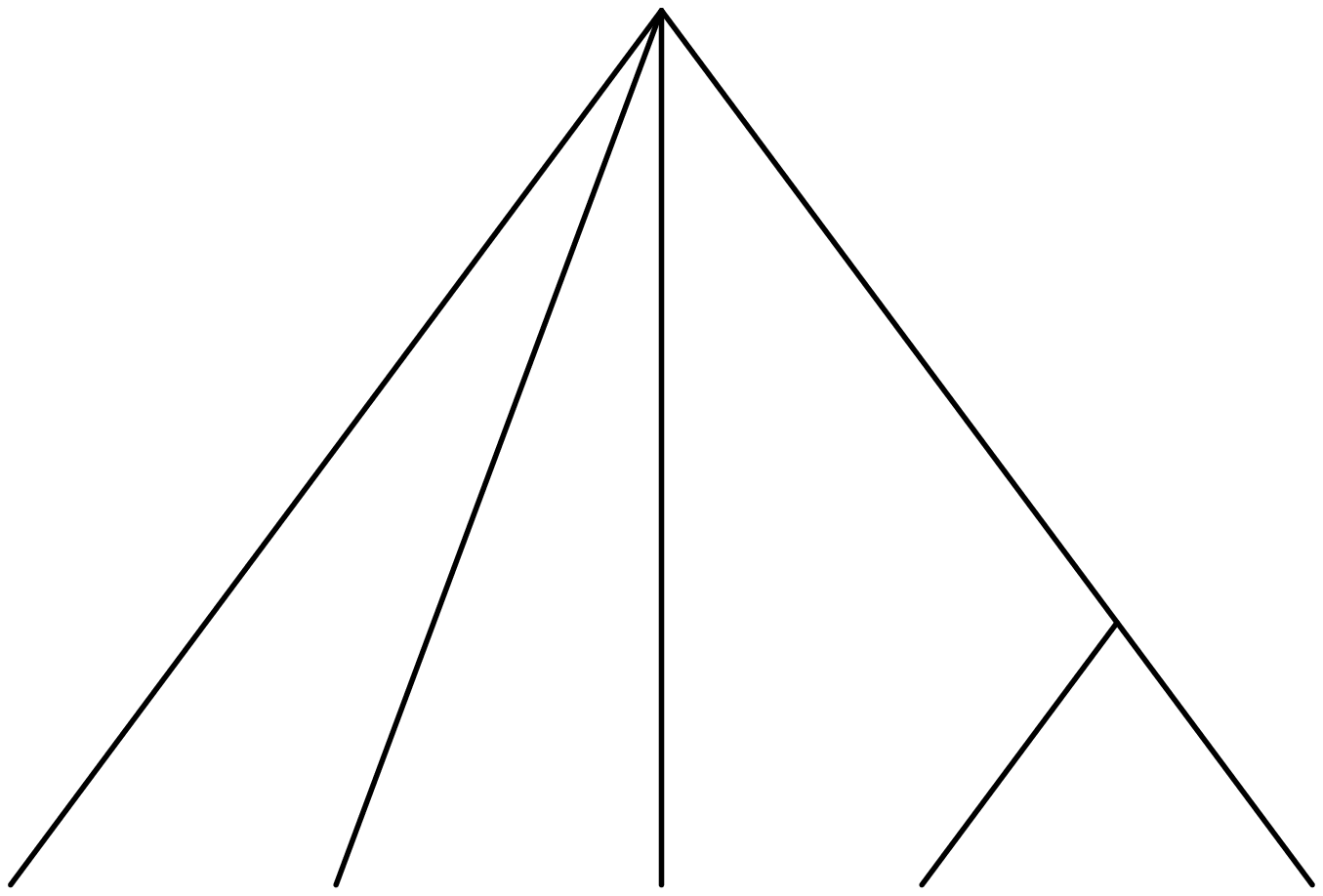} \\
$P_3$ & $((a,b, c,d)\tc z,e)$  & $u_3>u_1$ &  \includegraphics[width=.88cm]{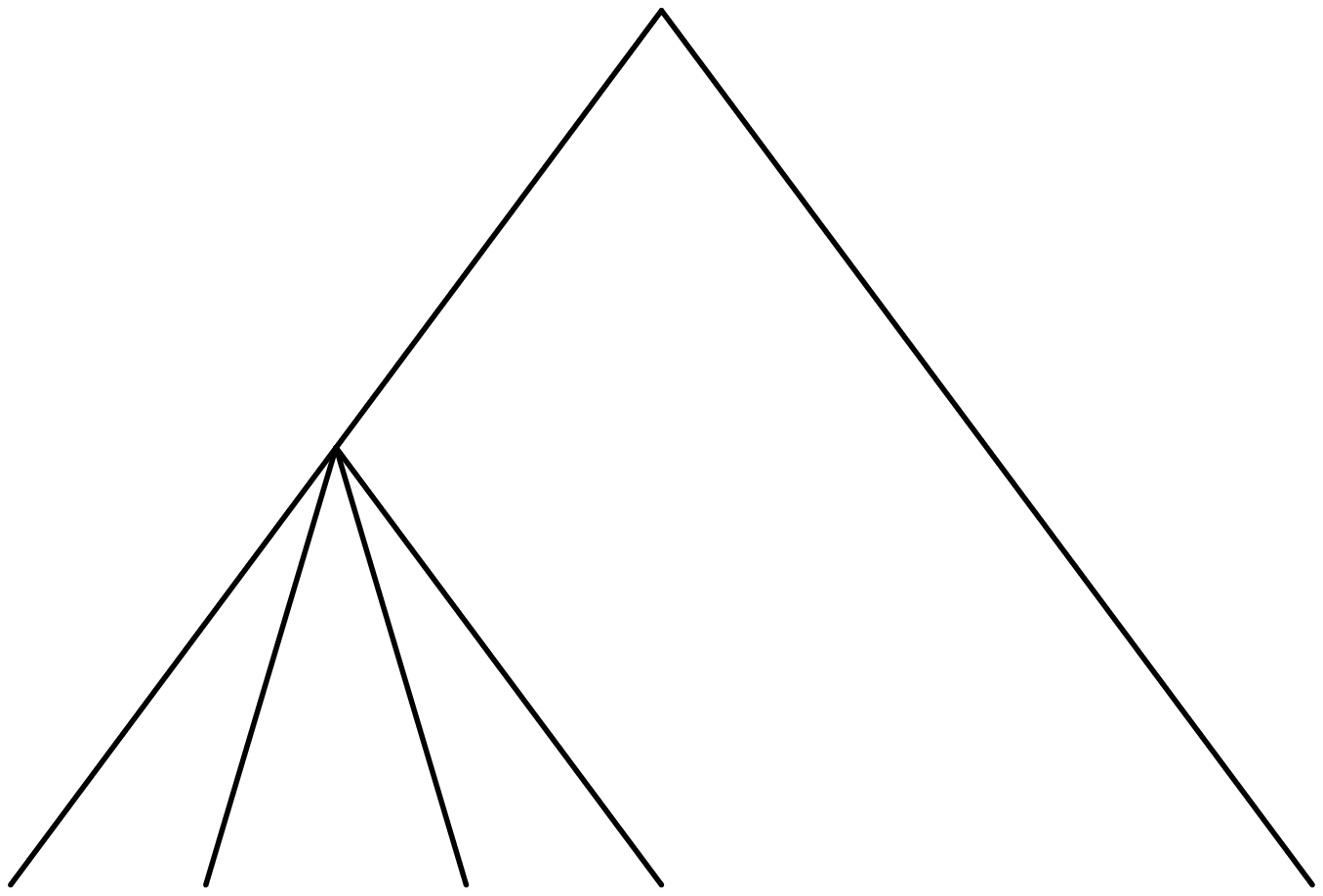}  \\
$P_4$ &  $((a,b,c)\tc y, d,e)$ & $u_1>u_2>u_7$ & \includegraphics[width=.88cm]{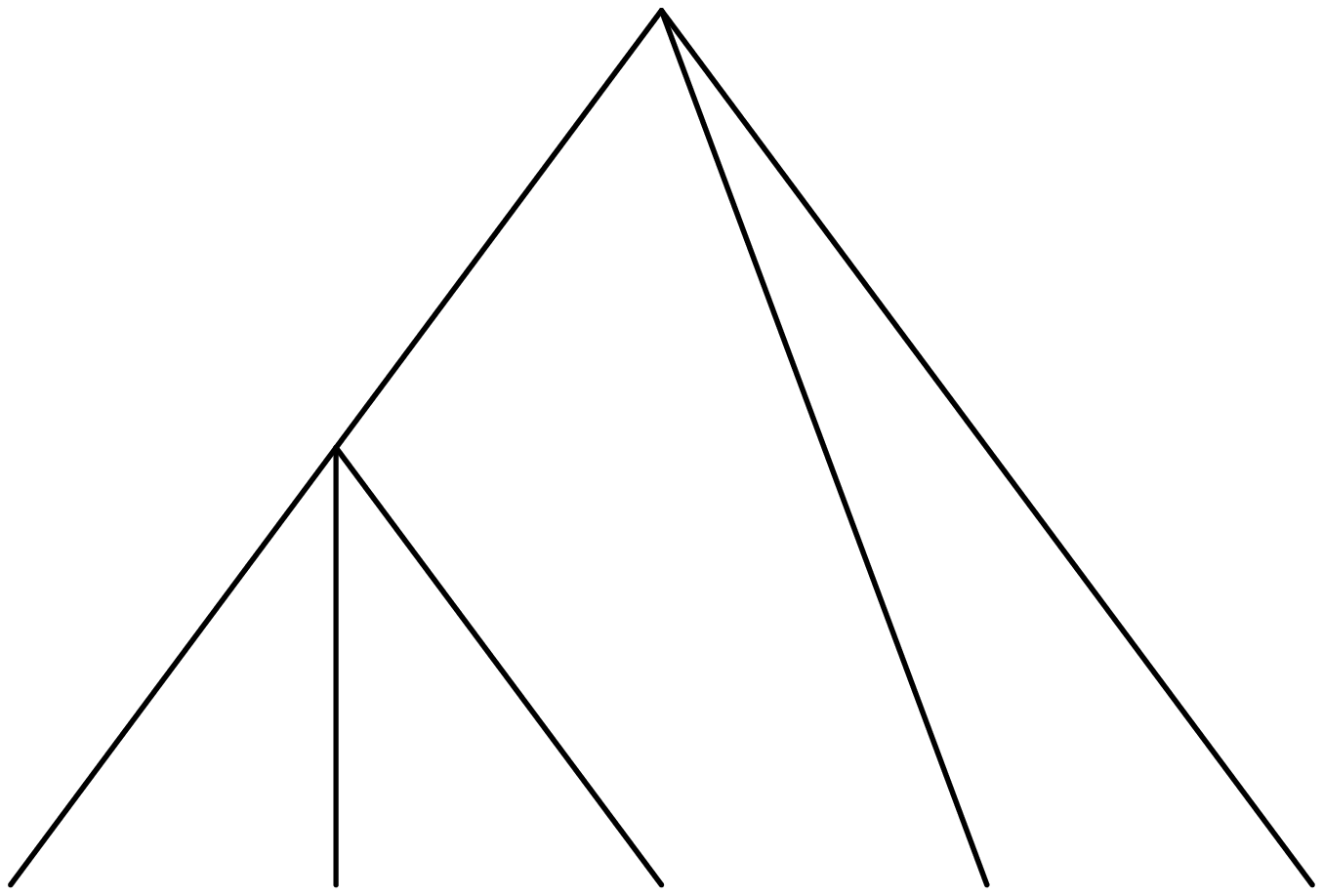}   \\
$P_5$ & $((a,b)\tc x, (d,e)\tc y, c)$  & $u_1 > u_2,u_4 > u_5$ &  \includegraphics[width=.88cm]{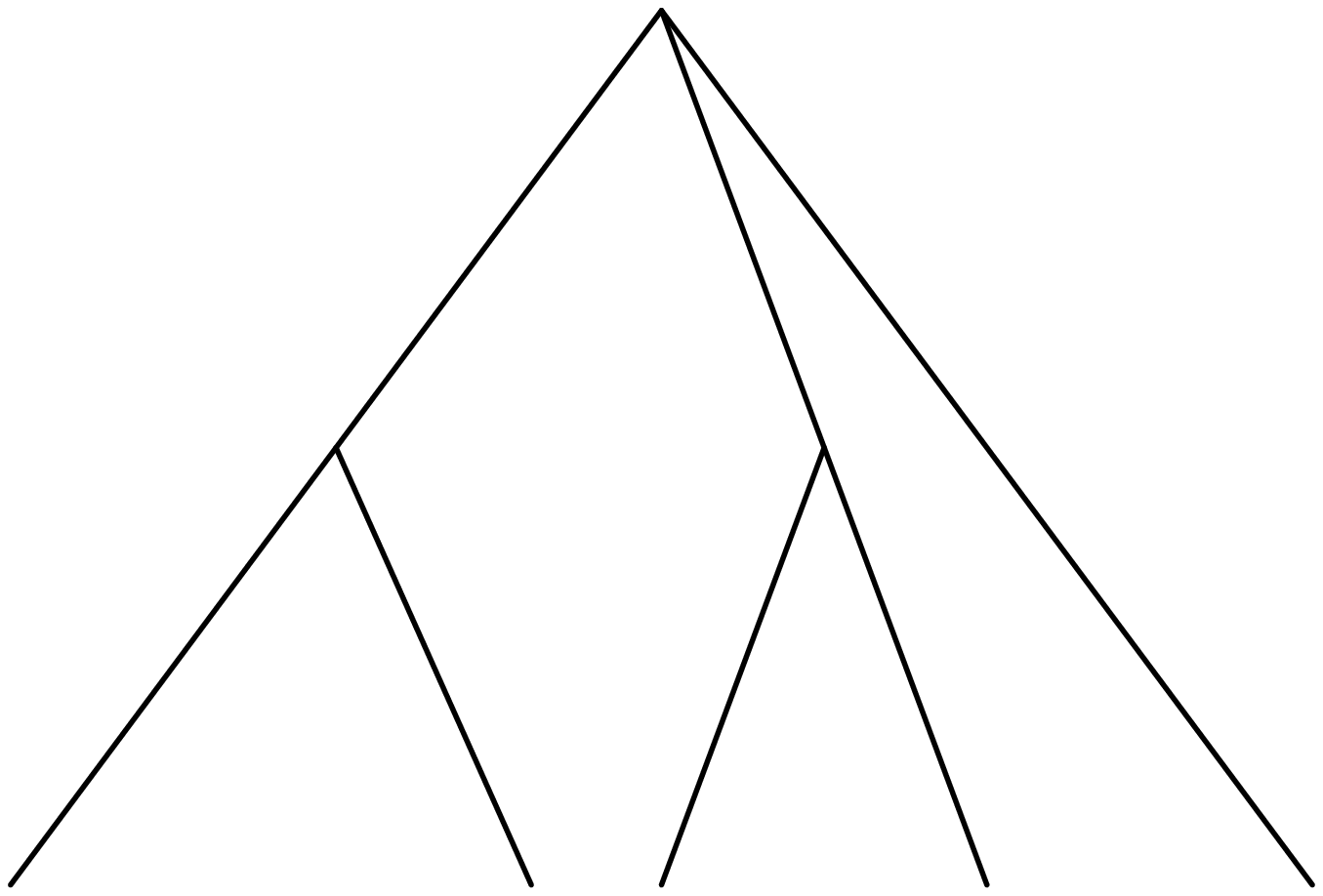}   \\
$P_6$  & $(((a,b)\tc x, c)\tc y, d, e)$  & $u_1 > u_2,u_4 > u_5 > u_7$ &    \includegraphics[width=.88cm]{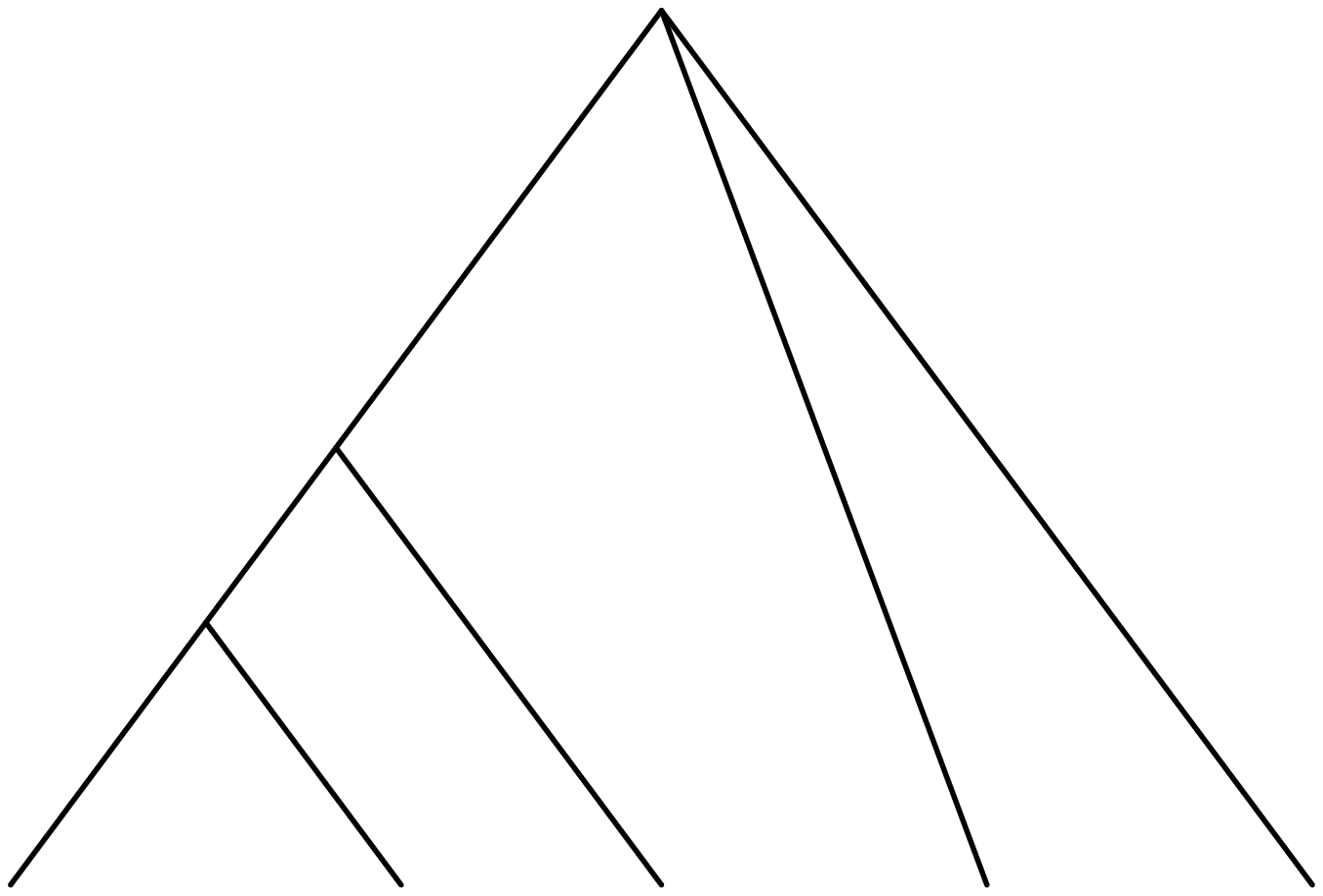} \\
$P_7$ &  $((a,b)\tc x, d,e)\tc z, c)$  & $u_1 > u_2, u_8 > u_4$ &    \includegraphics[width=.88cm]{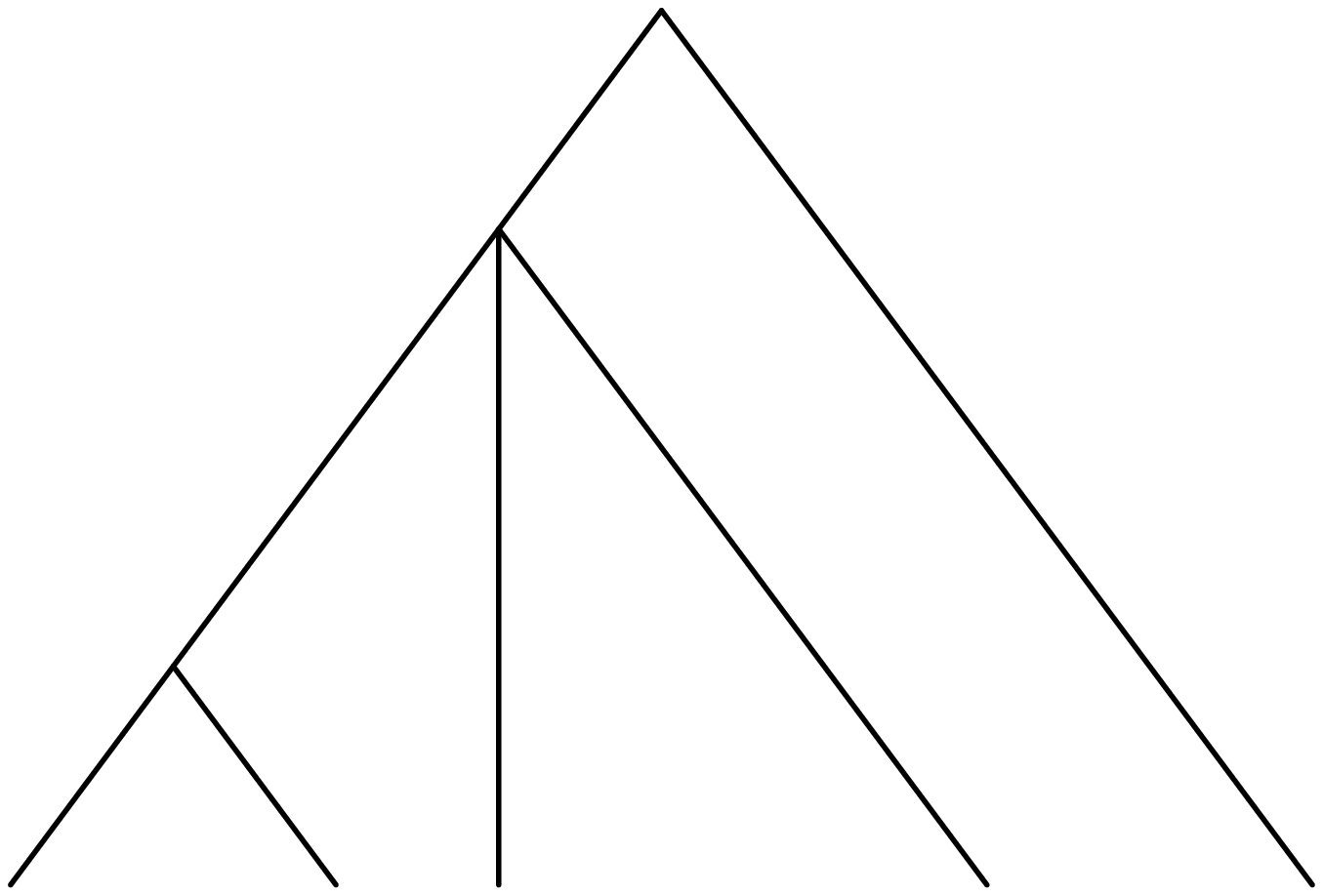}  \\
 $P_8$ & $(((a,b,c)\tc y, (d,e)\tc z)$ &  $u_1>u_2>u_7$  &    \includegraphics[width=.88cm]{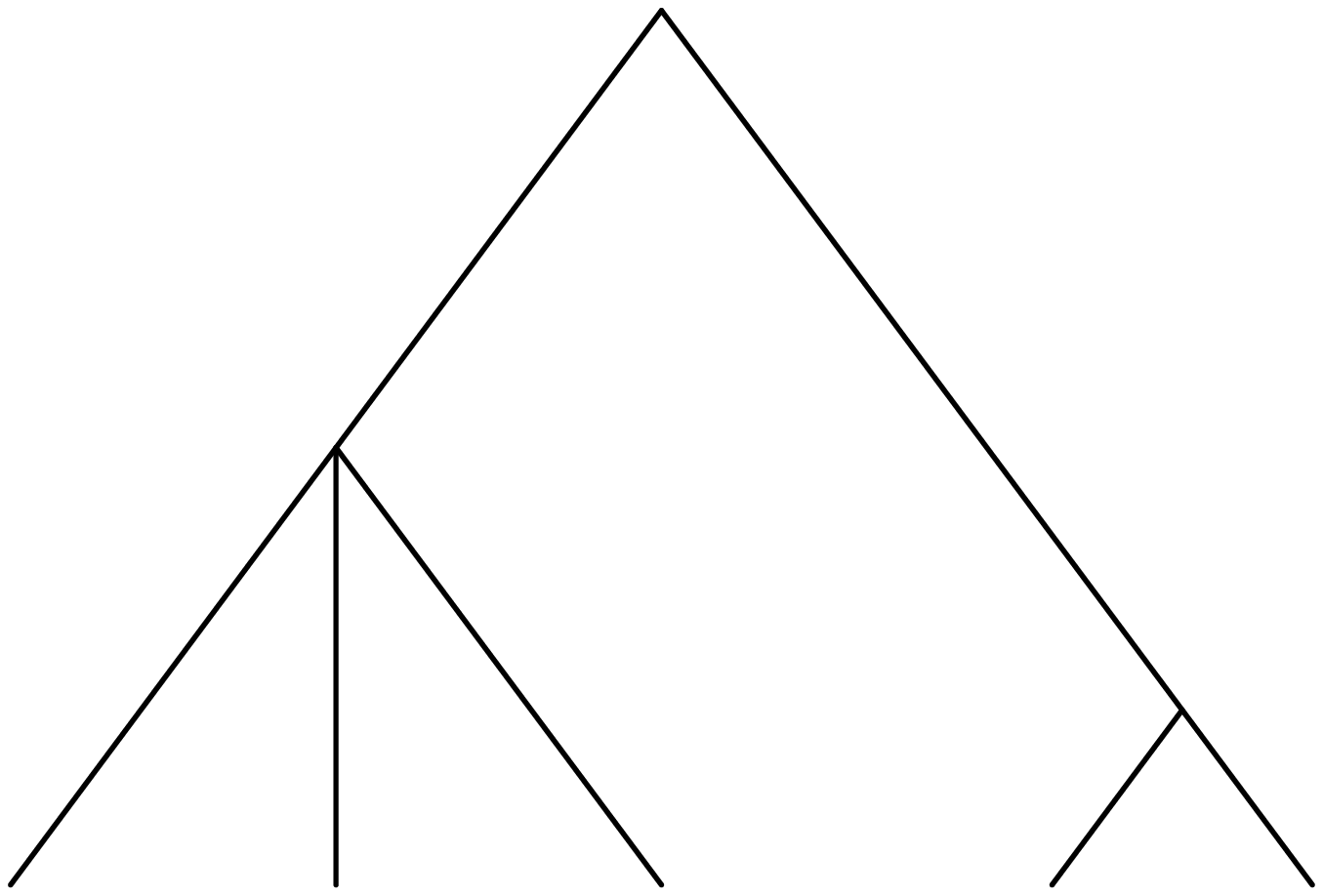} \\
$P_9$ & $(((a,b,c)\tc y, d)\tc z, e)$ & $u_1,u_3 > u_2 > u_7$ & \includegraphics[width=.88cm]{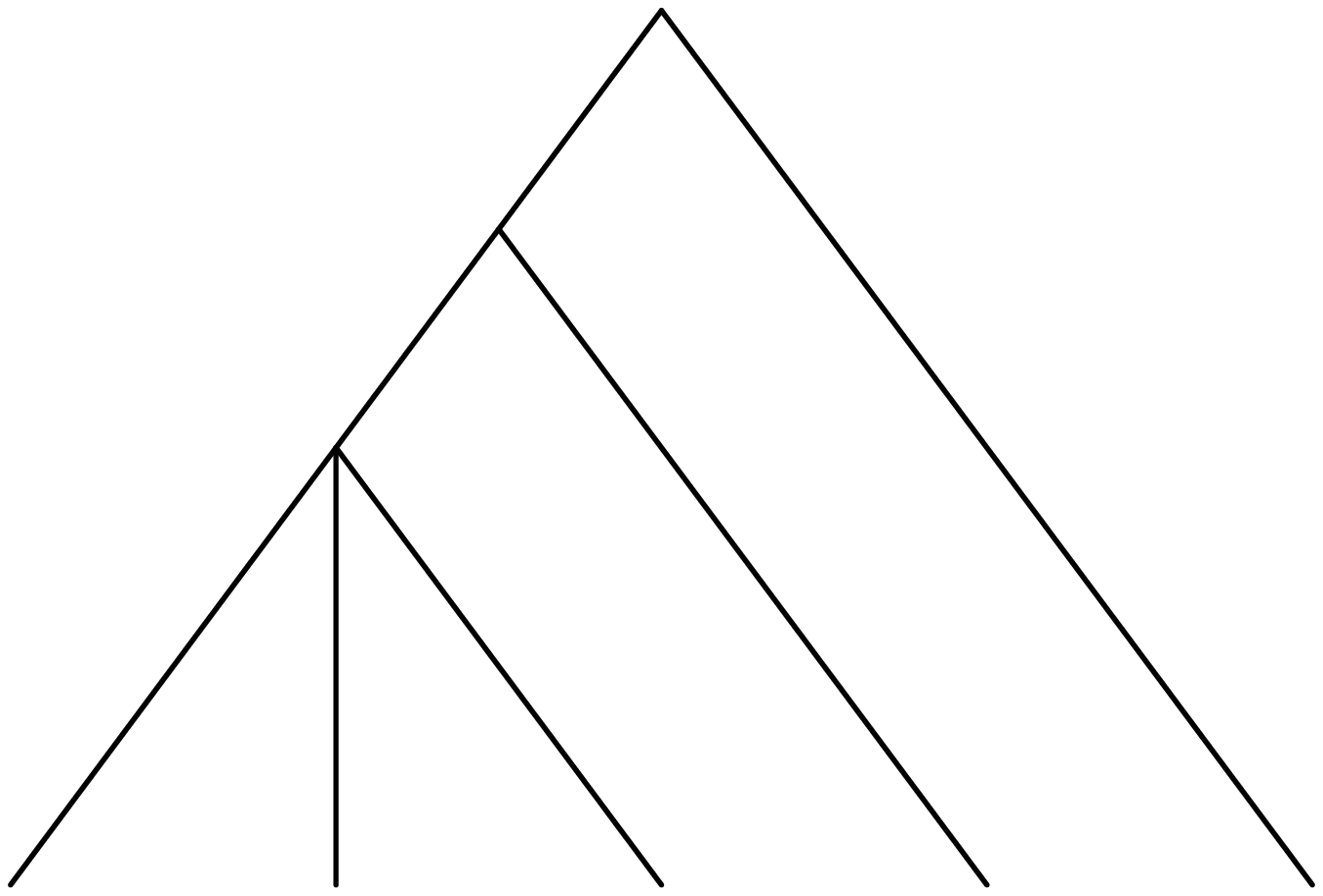}  \\[2ex]
 \hline
 \end{tabular}
 \end{center} 
\end{table}

\begin{table}
\begin{center}
\caption{Equivalence classes of equiprobable gene trees for each nonbinary 5-taxon species tree, and gene tree probabilities in terms of transformed branch lengths.
$\mathcal{T}$ denotes the set of all 5-taxon binary unrooted gene trees, $\{T_1, T_2, \dots, T_{15}\}$.}\label{T:poly2}
\begin{tabular}{l l l l}
 Species tree & equivalence classes & gene tree probabilities  \\[1.1ex]
\hline
 $P_1$ & $\mathcal{T} $ & $u_i = \frac{1}{15}$  &\\[1.1ex]
 $P_2$  & $\{T_1,T_4,T_{13}\}$ & $u_1 = \frac{1}{3} - \frac{4}{15}Z$ \\[1.1ex]
         & $\mathcal{T}\setminus \{T_1,T_4,T_{13}\}$ & $u_2 = \frac{1}{15}Z$ \\[1.1ex]
  $P_3$ 
           & $\{T_3,T_6,T_9\}$ & $u_3 = \frac{1}{9} - \frac{2}{45}Z^6$ \\[1.1ex]
           & $\mathcal{T}\setminus \{T_3,T_6,T_9\}$ & $u_1 = \frac{1}{18} + \frac{1}{90}Z^6$ \\[1.1ex]
$P_4$ & $\{T_1,T_4,T_{13}\}$ & $u_1 = \frac{1}{3} - \frac{1}{3}Y + \frac{1}{15}Y^3$ \\[1.1ex]
           & $\{T_2,T_3,T_5,T_6,T_9,T_{12}\}$ & $u_2 = \frac{1}{6}Y - \frac{1}{10}Y^3$\\[1.1ex]
           & $\{T_7,T_8,T_{10},T_{11},T_{14},T_{15}\}$ & $u_7 = \frac{1}{15}Y^3$\\[1.1ex]
 $P_5$ & $\{T_1\}$ & $u_1 = 1-\frac{2}{3}X-\frac{2}{3}Y + \frac{2}{5}XY$  \\[1.1ex]
 & $\{T_2,T_3\}$ & $u_2 = \frac{1}{3}Y - \frac{4}{15}XY$\\[1.1ex]
 & $\{T_4,T_{13}\}$ & $u_4 = \frac{1}{3}X - \frac{4}{15}XY$\\[1.1ex]
 & $\mathcal{T}\setminus\{T_1,T_2,T_3,T_4,T_{13}\}$& $u_5 = \frac{1}{15}XY$\\ [1.1ex]
 $P_6$ & $\{T_1\}$ & $u_1 = 1 - \frac{2}{3}X - \frac{2}{3}Y + \frac{1}{3}XY + \frac{1}{15}XY^3$ \\[1.1ex]
 & $\{T_2,T_3\}$ & $u_2 = \frac{1}{3}Y - \frac{1}{6}XY - \frac{1}{10}XY^3$\\[1.1ex]
 & $\{T_4,T_{13}\}$ & $u_4 = \frac{1}{3}X - \frac{1}{3}XY + \frac{1}{15}XY^3$\\[1.1ex]
 & $\{T_5,T_6,T_9,T_{12}\}$ & $u_5 = \frac{1}{6}XY - \frac{1}{10}XY^3$\\[1.1ex]
 & $\{T_7,T_8,T_{10},T_{11},T_{14},T_{15}\}$ & $u_7 = \frac{1}{15}XY^3$\\[1.1ex]
 $P_7$ & $\{T_1\}$ & $u_1 = \frac{1}{3} - \frac{2}{9}X - \frac{2}{45}XZ^6$\\[1.1ex]
 & $\{T_2,T_3\}$ & $u_2 = \frac{1}{3} - \frac{5}{18}X + \frac{1}{90}XZ^6$\\[1.1ex]
& $\{T_8,T_{11}\}$ & $u_8 = \frac{1}{9}X - \frac{2}{45}XZ^6$\\[1.1ex]
& $\mathcal{T}\setminus\{T_1,T_2,T_3,T_8,T_{11}\}$ & $u_4 = \frac{1}{18}X + \frac{1}{90}XZ^6$\\[1.1ex] 
$P_8$ & $\{T_1,T_4,T_{13}\}$ & $u_1 = \frac{1}{3} - \frac{1}{3}YZ + \frac{1}{15}Y^3Z$\\[1.1ex]
& $\{T_2,T_3,T_5,T_6,T_9,T_{12}\}$ & $u_2 = \frac{1}{6}YZ - \frac{1}{10}Y^3Z$\\[1.1ex]
& $\{T_7,T_8,T_{10},T_{11},T_{14},T_{15}\}$ & $u_7 = \frac{1}{15}Y^3Z$\\[1.1ex]
$P_9$ & $\{T_1,T_4,T_{13}\}$ & $u_1 = \frac{1}{3} - \frac{1}{3}Y + \frac{1}{18}Y^3 + \frac{1}{90}Y^3Z^6$\\[1.1ex]
& $\{T_2,T_5,T_{12}\}$ & $u_2 = \frac{1}{6}Y - \frac{1}{9}Y^3 + \frac{1}{90}Y^3Z^6$\\[1.1ex]
& $\{T_3,T_6,T_9\}$ & $u_3 = \frac{1}{6}Y - \frac{1}{18}Y^3 - \frac{2}{45}Y^3Z^6$\\[1.1ex]
& $\{T_7,T_8,T_{10},T_{11},T_{14},T_{15}\}$ & $u_7 = \frac{1}{18}Y^3 + \frac{1}{90}Y^3Z^6$\\[1.1ex]
\hline\\
\end{tabular}
\end{center}
\end{table}

 \end{document}